\documentclass[english,twocolumn,transaction]{IEEEtran}
\usepackage[T1]{fontenc}
\usepackage{float}
\usepackage{amsmath}
\usepackage{amsthm}
\usepackage{amssymb}
\usepackage{stackrel}
\usepackage{graphicx}
\usepackage{setspace}
\usepackage{url}
\usepackage{algorithm}
\usepackage{algorithmic}
\usepackage{booktabs}
\usepackage{cite}
\usepackage{color}
\usepackage{float}
\usepackage{subfigure}

\makeatletter

\floatstyle{ruled}
\newfloat{algorithm}{tbp}{loa}
\providecommand{\algorithmname}{Algorithm}
\floatname{algorithm}{\protect\algorithmname}

\theoremstyle{plain}

\theoremstyle{definition}

\theoremstyle{plain}

\theoremstyle{plain}

\theoremstyle{plain}

\IEEEoverridecommandlockouts
\usepackage{ifpdf}
\usepackage{cite}
\ifCLASSOPTIONcompsoc
\usepackage[caption=false,font=normalsize,labelfont=sf,textfont=sf]{subfig}
\else
\usepackage[caption=false,font=footnotesize]{subfig}
\fi
\usepackage{amsmath}
\@ifundefined{showcaptionsetup}{}{%
 \PassOptionsToPackage{caption=false}{subfig}}
\usepackage{subfig}
\makeatother

\usepackage{babel}
\newtheorem{define}{Definition}

\newtheorem{crit}{Criterion}
\newtheorem{remark}{Remark}
\renewcommand\figurename{Fig.}

\begin{document}
%
\title{Learning-Based Multi-Channel Access in 5G and Beyond Networks with Fast Time-Varying Channels}

\author{Shaoyang Wang,~\IEEEmembership{Student Member,~IEEE}, Tiejun Lv, \emph{Senior Member, IEEE},\\ Xuewei Zhang, Zhipeng Lin, and Pingmu Huang,

\thanks{Copyright (c) 2015 IEEE. Personal use of this material is permitted. However, permission to use this material for any other purposes must be obtained from the IEEE by sending a request to pubs-permissions@ieee.org. The financial support of the National Natural Science Foundation of China
(NSFC) (Grant No. 61671072) and BUPT Excellent Ph.D. Students Foundation (Grant No. CX2019106) are gratefully acknowledged.
(\emph{Corresponding author: Tiejun Lv.})

S. Wang, T. Lv, X. Zhang, Z. Lin and P. Huang are with the School of Information and Communication Engineering, Beijing University of Posts and Telecommunications (BUPT), Beijing 100876, China (e-mail: \{shaoyangwang, lvtiejun, zhangxw, linlzp, pmhuang\}@bupt.edu.cn). Z. Lin is also with the School of Electrical and Data Engineering, University of Technology Sydney (UTS), Sydney 2007, Australia.
}}
\maketitle

\begin{abstract}
We propose a learning-based scheme to investigate the dynamic multi-channel access (DMCA) problem in the fifth generation (5G) and beyond networks with fast time-varying channels wherein the channel parameters are unknown. The proposed learning-based scheme can maintain near-optimal performance for a long time, even in the sharp changing channels. This scheme greatly reduces processing delay, and effectively alleviates the error due to decision lag, which is cased by the non-immediacy of the information acquisition and processing. We first propose a psychology-based personalized quality of service model after introducing the network model with unknown channel parameters and the streaming model. Then, two access criteria are presented for the living streaming model and the buffered streaming model. Their corresponding optimization problems are also formulated. The optimization problems are solved by learning-based DMCA scheme, which combines the recurrent neural network with deep reinforcement learning. In the learning-based DMCA scheme, the agent mainly invokes the proposed prediction-based deep deterministic policy gradient algorithm as the learning algorithm. As a novel technical paradigm, our scheme has strong universality, since it can be easily extended to solve other problems in wireless communications. The real channel data-based simulation results validate that the performance of the learning-based scheme approaches that derived from the exhaustive search when making a decision at each time-slot, and is superior to the exhaustive search method when making a decision at every few time-slots.
\end{abstract}


\section{Introduction}
\renewcommand\figurename{Fig.}
With the popularity of applications in high-speed railways and highways, high-mobility environments attract great attention in the fifth generation (5G) and beyond mobile communications, leading to many new challenges, such as fast time-varying channels, frequent handovers, and complex channel environments \cite{8626080}. At the same time, it is an important ability of 5G and beyond networks to efficiently support ultra-reliable and low-latency communication (URLLC) services for autonomous driving and vehicle-to-everything (V2X) scenarios\cite{8004168}. Providing stable, reliable, and fast data transmission service is also a crucial target in high-mobility wireless communications. As a most important component in achieving the above targets \cite{7405718}, dynamic multi-channel access (DMCA), should be greatly improved \cite{8254980}, since it can also effectively promote system throughputs \cite{4570228}.

Typical DMCA causes mismatch between user requirements and channel supplies, because transmitter probes the idle channels before accessing, and channels are randomly selected. Although various optimization schemes, e.g., myopic policy and its variants, have been proposed \cite{6230480,6818431,5593977,5398950,8085106,7029097}, they only consider low-mobility scenarios, and are based on Markov model with known transition probabilities, orders, and etc.

The high-mobility scenarios pose two severe challenges. Firstly, the optimal choice in the time-slot $t$ may become the worst choice in the next time-slot $t\!+\!1$ due to sharp channel changes. The most intuitive alternative is to make the optimal choice at every time-slot; however, this leads to poor system stability, and increases the processing complexity and service delay. Meanwhile, frequent handovers and group handovers might increase the drop-off probability, leading to the degradation of the quality of service (QoS) \cite{8081737}. Secondly, in practice, a DMCA strategy, which is employed at time-slot $t$, will actually executed at time-slot $t\!+\!\tau$, due to the inevitable delay $\tau$ caused by the information acquisition and processing.  In this paper, this delay is defined as \emph{non-instant decision error}. However, it is generally neglected in the literature. In addition, the parameters of the Markov channel model are difficult to be obtained in practice.

The difference in service demand is universal in human society. Regrettably, from the first generation to the forth generation communications, the individual requirements of users have not been met, let alone the \emph{subjective experience} \cite{8603730}. For instance, for the same service rate, some users subjectively feel fast, but other users feel slow. For this issue, in 5G and beyond mobile communications, the \emph{subjective experience} needs to be analyzed and modeled quantitatively. Besides differential requirements of individual users, the requirements of same users are also be different over the time horizon. Therefore, we need to consider personalized QoS (PQoS), which further increases the challenges of DMCA design.

Recently, the rapid development of the artificial intelligence and machine learning has attracted much attention from the industry and academia \cite{huang2019deep,8233654}. Machine/deep learning has been applied to physical layer technology implementation, and remarkable performance has been achieved in many scenarios, e.g., super-resolution channel estimation and direction of arrivals estimation \cite{8400482}, hybrid pre-coding under millimeter-wave massive multiple-input-multiple-output system \cite{8618345}, non-orthogonal multiple access system \cite{8454272,8387468}, beamforming design \cite{8880526}, modulation recognition \cite{8645696}, wireless scheduling \cite{8664604}, and etc. Different from the above applied deep learning technologies, reinforcement learning (RL) mainly urges the agents to maximize the discounted reward occurred over a finite time horizon and discover the optimal strategy by their own. RL-based algorithms have been proposed for several communication scenarios, e.g., dynamic channel selection \cite{6522954}, aggregated interference control \cite{5415567}, inter-cell interference coordination \cite{6965655} and resource allocation \cite{8370091}. The main distinctions of these RL algorithms are reflected in the variations of Q-learning. They are all based on a look-up table storing the value of state-action pair, which is learned by using either temporal-difference or fuzzy rules. When the action/state space increase, it will be difficult to store and look up the Q table in the Q-learning.

Deep reinforcement learning (DRL) integrates the advantages of deep learning into RL, and it can overcome the difficulties in the Q-learning. There are some efforts devoted to combining DRL with the DMCA or spectrum access \cite{8303773,8254101,8532121,8610402}. Nevertheless, the adopted DRL technique, i.e., the deep Q network (DQN) \cite{8303773,8254101,8532121,8610402}, can only tackle a very limited action spaces, which is not applicable to the complicated DMCA due to high-dimensional/continuous action outputs in high-mobility communications. Furthermore, DRL still has convergence problems in high-mobility environments, although it usually converges faster than RL \cite{8665952}.

Due to lack of design paradigm, applying DRL to wireless communications is still difficult. Considering the complicated variations in communication scenarios, we believe it is worth further studying the differences between the communication application scenarios and the traditional DRL-applied scenarios. Directly applying the DRL to solve communication problems, as in \cite{6522954,5415567,6965655,8370091,watkins1992q,8303773,8254101,8532121,8610402,8665952}, might not be the best solution, especially in high-mobility environments.

Against the above backdrops, in this paper we investigate the DMCA design in high-mobility environments with unknown channel parameters by using a learning-based scheme. The major contributions are summarized as follows:
\begin{itemize}
\item The \emph{non-instant decision error} and the \emph{subjective experience} are considered for the first time in the DMCA problem in fast time-varying channels. We propose a psychology-based PQoS (P-PQoS) principle. It is a portrayal for the \emph{subjective experience} of users about service delay.
\item A novel DRL algorithm, namely, prediction-based deep deterministic policy gradient (P-DDPG), is proposed to address the difficulties of the high-dimensional action/state spaces and the slow convergence speed. The real channel data-based prediction results confirm the channel predictability under specific scenarios, which may have some guiding significance for the estimation/modeling of fast time-varying channel. The channel parameters do not need to be known in our scheme.
\item A learning-based DMCA framework is designed exploiting the characteristics of high-mobility system, and the corresponding parameter configurations and recommendations are also given. The framework can sense future channel changes and determine the strategy to ensure high-performance for a period of time, and thus it is able to effectively reduce the processing delay, and alleviate the impact of the \emph{non-instant decision error}. The learning-based DMCA framework consists of two modules: a channel prediction module (CPM) responsible for the channel prediction and a P-DDPG module (leverages the P-DDPG algorithm to learn and update) responsible for outputting final channel access policies.
\item The ideas of virtual user and the design paradigm of the reward function are proposed, which are used to ensure the stability of the DMCA network framework and further improve the convergence speed, respectively. In addition, the specific design paradigm of solving the communication problems with DRL is also introduced and demonstrated. This paradigm profoundly exploits the characteristics of DMCA and exhibits strong universality.
\end{itemize}

The rest of the paper is organized as follows. In Section \uppercase\expandafter{\romannumeral2}, we first introduce the DMCA network model. Then, two streaming models are discussed, andthe definition of P-PQoS is given. In Section \uppercase\expandafter{\romannumeral3}, two access criteria are described, and the optimization problems are formulated. In Section \uppercase\expandafter{\romannumeral4}, the DDPG algorithm is first introduced, and then a novel P-DDPG algorithm, as the key algorithm of the DMCA implementation, is proposed. In Section \uppercase\expandafter{\romannumeral5}, the overall implementation of the learning-based DMCA framework is introduced, and the corresponding CPM module and P-DDPG module are also elaborated. In Section \uppercase\expandafter{\romannumeral6}, we present simulation results and analyses, and this paper is concluded in Section \uppercase\expandafter{\romannumeral7}.
\section{System Model}
\subsection{Network Model}
Consider a classical time-slotted multi-channel network where a single-antenna base station (BS) serves $N$ single-antenna high-mobility users (denoted as $\mathcal{N}=\{1,2,\cdots,N\}$) simultaneously through $M$ (denoted as $\mathcal{M}=\{1,2,\cdots,M\}$, $M\!\geq\!N$) channels. $N$ is a variable due to user mobility.

Currently, the existing related researches are focusing on the first-order ($k$=1) two-state model\footnote{The two-state means $\left\{ \mathrm{good},\mathrm{bad}\right\}$. If the information is successfully transmitted, the state is $\mathrm{good}$, otherwise, the state is $\mathrm{bad}$.} \cite{4723352,8303773} or the $X$-state \cite{7744533} model under the premise that $\mathbb{P}\left[\cdot|\cdot\right]$ is known, where $\mathbb{P}\left[\cdot|\cdot\right]$ represents the conditional probability and $k$ is the order of the Markov model. The existing Markov models for multi-channel access require some parameters, including state transition probability, the order of the Markov model, and the number of channel states, etc. These parameters are difficult to obtain in real-word scenarios. In this paper, we avoid the traditional Markov modeling for the wireless communication channels. As a result, the channel transition probability $\mathbb{P}\left[\cdot|\cdot\right]$, the order $k$ and the number of states do not need to be known.

In this paper, two access cases are considered, as given by
\begin{itemize}
  \item \emph{Single-user access}: At the beginning of time-slot $t$, the user selects one channel from $M$ channels to access and transmit messages. The transmit rate\footnote{The transmission rate is a function of time $t$, since the fast time-varying channels are considered.} of the channel $m$ is defined as
\begin{align}
R_{m}(t)=B_{m}\log\left(1+\frac{|h_{m}(t)|^{2}P_{m}}{\sigma_{m}^{2}}\right),
\end{align}
where $h_{m}\left(t\right)$ is the gain of the channel $m$ in the time-slot $t$; $P_{m}$ is the power of channel $m$; and $\sigma_{m}^{2}$ is the noise power of the channel $m$. $B_{m}$ is the bandwidth of channel $m$ and $B_{m}\!=\!B/M$ since we assume that the frequency band $B$ is equally allocated to $M$ channels.
  \item \emph{Multi-user access}: At the beginning of each time-slot, $N$ users select $N$ channels from $M$ channels to access and transmit messages.
\end{itemize}
\subsection{Streaming Model}
Two data traffic models are introduced as follows:
\subsubsection{Living Streaming Model (LSM)}
The LSM focuses on the voice call, live television broadcasts, etc. In LSM, the traffic data cannot be cached in advance and can only be obtained in real-time.
\subsubsection{Buffered Streaming Model (BSM)}
The BSM does not have the real-time requirements, such as downloading services, multimedia video services and so on. To this end, the traffic data in BSM can be cached in advance.
\subsection{User Request Model}
We then proceed to the user request model.  Firstly, the concept of \emph{delay sensitivity} is proposed.
\begin{define}
\emph{Delay sensitivity:} This concept depicts the difference in psychological perception between different users for same delay. The delay sensitivity of user $n$ is defined as
\begin{align}
\lambda_{n}=\frac{1}{2}-\frac{1}{\pi}\arctan\left(\tau_{\mathrm{limit},n}-\tau_{\mathrm{real}}\right),
\end{align}
where $\lambda_{n}\in\left(0,1\right),\forall n$; $\tau_{\mathrm{real}}$ is the standard delay value specified in the white paper \cite{baipishu}; $\tau_{\mathrm{limit},n}$\footnote{In the actual deployment, in order to obtain $\tau_{\mathrm{limit},n}$, we need to perform statistical analysis on user behavior according to big data technology, obtain different estimation values, and then continuously adjust this parameter by the online method. Taking into account the scope of this paper, we assume $\lambda_{n}$ is known.} is a limiting value that can be tolerated by user $n$.
\end{define}

Larger the value of $\left|\tau_{\mathrm{limit},n}-\tau_{\mathrm{real}}\right|$ is, the less obvious the effect of the delay on the subjective feelings of users \cite{8485468} is, which can be well described by $\arctan(x)$. Based on Definition 1, we propose the P-PQoS constraint, which can truly meet the personalized services requirements and is the trend for the future development. P-PQoS is defined as
\begin{align}\label{P-PQoS}
R_{\mathrm{user},n}(t)=(2\lambda_{n}+\beta(1-2\lambda_{n}))R_{\mathrm{user}}(t),
\end{align}
where $R_{\mathrm{user}}(t)$ is the minimum data rate constraint for a certain service; $\beta$ is an influence factor of the \emph{subjective experience}. Larger $\beta$ indicates smaller influence of the \emph{subjective experience}. (\ref{P-PQoS}) indicates the difference of the subjective expected service rates between different users for the same service.

\section{Problem Formulation}
To deploy the  above communication systems, at each time-slot, we select $N$ channels, observe their states, and use them to transmit messages. Let $R_{m}(t)$ denote the rate state of the channel $m$ at time-slot $t$, then we have the state vector $\mathbf{S}_{\mathrm{ch}}(t)=\left[R_{1}(t),R_{2}(t),\cdots,R_{N}(t)\right]$. Let $\mathbf{A}(t)=\left[\mathcal{A}_{1}(t),\mathcal{A}_{2}(t),\cdots,\mathcal{A}_{N}(t)\right]$ denote the decision vector at time-slot $t$, where $\mathcal{A}_{n}(t)\in\mathcal{M}$ and $\mathcal{A}_{i}(t)\neq\mathcal{A}_{j}(t)$ for $\forall i,j\in\mathcal{N}$.

\begin{define}
\emph{P-PQoS bias}: Let $\mathbf{w}(t)$ denote the difference between the P-PQoS and the service rates provided by channels in the time-slot $t$. The P-PQoS bias within the $T$ time periods can then be expressed as
\begin{align}
Q(T)\triangleq\left\{ \mathbf{w}(1),\cdots,\mathbf{w}(T)\right\},
\end{align}
where
\begin{align}
&\mathbf{w}(t)=\left[\varDelta_{1}(t),\cdots,\varDelta_{N}(t)\right],\nonumber \\
&\varDelta_{n}(t)\triangleq R_{\mathcal{A}_{n}(t)}(t)-R_{\mathrm{user},n}(t).\nonumber
\end{align}
\end{define}

\begin{define}
\emph{Non-instant decision error}: Let $\rho(\mathbf{S}(t))$ denote the performance of system\footnote{The system performance can be evaluated by the system throughput in this paper, of course, it can also be measured by the system stability, energy consumption, complexity, and so on.} when making decision $\mathbf{A}(t)$ based on state $\mathbf{S}(t)$, and let $\varDelta t$ represent the time difference between observation and decision. The performance error caused by delay $\varDelta t$ can be defined as
\begin{align}
\Omega(\varDelta t)=|\rho(\mathbf{S}(t))-\rho(\mathbf{S}(t+\varDelta t))|/\rho(\mathbf{S}(t+\varDelta t)).
\end{align}
Note that $\mathbf{S}(t)$ also includes other state information in addition to $\mathbf{S}_{\mathrm{ch}}(t)$, which will be analyzed in detail later.
\end{define}

In LSM,  it is a resource waste when the provided service rate $R_m(t)$ is higher than the requested service rate $R_{\mathrm{user},n}(t)$. For this issue, we have the following criterion.
\begin{crit}
In LSM, the optimal strategy for user $n$ can be expressed as
\begin{align}
\ensuremath{\mathcal{A}_{n}(t)}\triangleq\mathcal{I}(\min\{\varDelta_{1}(t),\cdots,\varDelta_{N}(t)|\varDelta_{n}(t)\geq0\}),
\end{align}
where $\mathcal{I}(x)$ is an index function. For instance, $\mathcal{I}(\varDelta_{n}(t))=n$.
\end{crit}

Criterion 1 is called the \emph{minimization after satisfaction} criterion. Therefore, the DMCA problem in LSM can be described as
\begin{subequations}
\begin{align}
\underset{\mathbf{A}(t)}{\mathrm{argmin}}&\stackrel[n=1]{N}{\sum}\left(R_{\mathcal{A}_{n}(t)}(t)-R_{\mathrm{user},n}(t)\right), \tag{\emph{OP1}} \\ \nonumber
\mbox{s.t.} \quad
&\varDelta_{n}(t)\geq0,\forall n. \nonumber
\end{align}
\end{subequations}

Unlike LSM, the user in BSM always expects that the provided service rate is as large as possible. In this case, the optimal strategy is described in criterion 2.
\begin{crit}
In BSM, the optimal strategy for user $n$ can be expressed as
\begin{align}
\ensuremath{\mathcal{A}_{n}(t)}\triangleq\mathcal{I}(\max\{\varDelta_{1}(t),\cdots,\varDelta_{N}(t)|\varDelta_{n}(t)\geq0\}).
\end{align}
\end{crit}

Criterion 2 is called the \emph{maximization after satisfaction} criterion. Thus, the DMCA problem for BSM can be formulated as
\begin{subequations}
\begin{align}
\underset{\mathbf{A}(t)}{\mathrm{argmax}}&\stackrel[n=1]{N}{\sum}\left(R_{\mathcal{A}_{n}(t)}(t)-R_{\mathrm{user},n}(t)\right), \tag{\emph{OP2}} \\ \nonumber
\mbox{s.t.} \quad
&\varDelta_{n}(t)\geq0,\forall n. \nonumber
\end{align}
\end{subequations}

It is noted that a user will be disconnected when the cache space is full, and then, a new user will be admitted to access, due to the limited cache space in BSM.

In fact, considering the variability of channel states and the diversity of user requirements, it is not always possible to find $N$ channels that can satisfy P-PQoS simultaneously, i.e., $\varDelta_{n}(t)\geq0$ for $\forall n$. To this end, let $\vartheta(t)$ represent the \emph{service success rate}, denoted as
\begin{align}
\vartheta(t)=\left(\underset{n\in\mathcal{N}}{\sum}\left(\varGamma\left(\varDelta_{n}\left(t\right)\geq0\right)\right)\right)/N.
\end{align}
If $x$ is true, $\varGamma(x)=1$ is 1; otherwise, $\varGamma(x)=0$.

In this paper, we always give the priority to maximum the number of successfully served users. Thus, the optimization problems $OP1$ and $OP2$ can be further expressed as a two-step optimization problems:
\begin{itemize}
  \item \emph{Step 1}:
\begin{align}\label{step1}
\Theta(t)=\underset{\mathbf{A}(t)}{\mathrm{argmax}}\:\vartheta(t).
\end{align}
  \item \emph{Step 2}:
\begin{align}\label{step21}
OP1:\quad\underset{\mathbf{A}(t)\subseteq\Theta(t)}{\mathrm{argmin}}\underset{n\in\mathcal{N}}{\sum}\left|R_{\mathcal{A}_{n}(t)}(t)-R_{\mathrm{user},n}(t)\right|,
\end{align}
or
\begin{align}\label{step22}
OP2:\quad\underset{\mathbf{A}(t)\subseteq\Theta(t)}{\mathrm{argmax}}\underset{n\in\mathcal{N}}{\sum}\left(R_{\mathcal{A}_{n}(t)}(t)-R_{\mathrm{user},n}(t)\right).
\end{align}
\end{itemize}

Traditional optimization method, such as convex optimization, can hardly be tailored for the above problems since the problem is highly flexible and the dynamic channel information is unknown. In addition, $\Omega(\varDelta t)$ will become more noticeable when meeting fast time-varying channels. To solve the problems, we propose a new approach. We first propose the P-DDPG algorithm, and then, based on this algorithm, a novel learning-based DMCA framework is implemented to obtain the dynamic channel access policies.

\section{Proposed P-DDPG Algorithm}
Generally speaking, when the dynamic characteristics of the system are unknown, there are two main approaches to solving these problems:
\begin{itemize}
  \item \emph{Model-based approach}: It first estimates the system model from observations, and then the dynamic programming (DP) or heuristic policy is adopted, such as myopic policy or whittle index policy \cite{5208571,5605371}.
  \item \emph{Model- free approach}: It learns the policy directly through interactions with the system without estimating the system model.
\end{itemize}

The model-based approach is less preferred since its limited observation capability may result in a bad model estimation. We also note that, even if the system dynamic is well estimated, solving a partially observable Markov decision process in a large state space is always a bottleneck as the DP method has exponential time complexity and the heuristic approaches cannot have any performance guarantee. All these challenges motivate us to apply the model-free method. By incorporating the idea of RL, this method can learn directly from observations without the necessity of finding an estimated system model, which can be easily extended to complicated systems.

Although irreducible, $\Omega(\varDelta t)$ can be weaken as much as possible. Specifically, the strategy is adopted at every time-slot in the traditional method, and thus the impact of $\Omega(\varDelta t)$ is $T$ times in $T$ periods. If a strategy can maintain the high-performance over a time period as the system changes, such as performing well in $5$ time-slots, the strategy only needs to be adopted at time $t=1,6,11,...$ In this case, the impact of $\Omega(\varDelta t)$ can be reduced to $T/5$ of the previous one. The RL pursues the long-term reward of agents, which can effectively solve the \emph{non-instant decision error} problem.

In addition, by employing RL, the execution time of the algorithm can be broadened, and the requirement of computing power for the terminal equipment can be also reduced, because there is no need to adopt a strategy at every time-slot $t$.

\subsection{Overview of RL}
The model-free RL algorithms contain the value-based and the policy-based ones. The Q-learning, Sarsa and DQN \cite{mnih2015human} are value-based, and the DDPG \cite{lillicrap2015continuous} is policy-based. The main advantage of the policy-based RL algorithm is that it is effective to confront with the challenges caused by continuous or high-dimensional action spaces \cite{silver2014deterministic}. The DDPG is a powerful algorithm combining the advantages of DQN and Actor-Critic (AC) structure \cite{bhatnagar2009natural}.

\begin{figure}[ht]
\centering{}\includegraphics[scale=0.47]{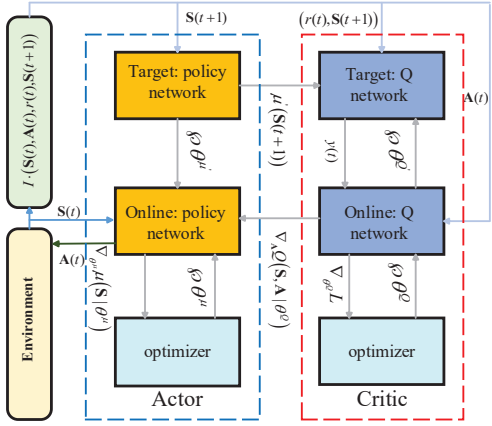}
\caption{The flow diagram of DDPG algorithm. The green box means the ``experience replay'' pool and $\wp$ means the network parameter update.}\label{DDPG}
\end{figure}

The principle of DDPG algorithm is shown in Fig. \ref{DDPG}. The AC architecture consists of a critic network, also known as the Q network, and an actor network, also known as the policy network. The DDPG creates two copies for the critic network and the actor network, called the online network and the target network, respectively. This method can improve the stability and accelerate the convergence speed.

The action value function $Q^{\mu}\left(\mathbf{S}(t),\mathbf{A}(t)\right)$ is used to represent the reward of executing action $\mathbf{A}(t)$ according to strategy $\mu$ when the state is $\mathbf{S}(t)$. The specific formula of $Q^{\mu}\left(\mathbf{S}(t),\mathbf{A}(t)\right)$ is defined as
\begin{align}\label{Q_update}
Q^{\mu}\left(\mathbf{S}(t),\mathbf{A}(t)\right)=&\mathbb{E}_{r_{t},\mathbf{S}(t+1)\sim E}[r(\mathbf{S}(t),\mathbf{A}(t))\nonumber\\
+&\gamma Q^{\mu}\left(\mathbf{S}(t+1),\mu\left(\mathbf{S}(t+1)\right)\right)], \end{align}
where $r(\mathbf{S}(t),\mathbf{A}(t))$ denotes the immediate reward, and $\gamma\in\left(0,1\right]$ is a discount factor.

The gradient of the policy network is given by
\begin{align}\label{Policy_loss}
\nabla_{\theta^{\mu}}\mu\approx\mathbb{E}_{\mu^{'}}&\left[\nabla_{\theta^{\mu}}Q\left(\mathbf{S},\mathbf{A}|\theta^{Q}\right)|_{\mathbf{S}
=\mathbf{S}(t),\mathbf{A}=\mu\left(\mathbf{S}(t)|\theta^{\mu}\right)}\right] \nonumber\\
=\mathbb{E}_{\mu^{'}}&[\nabla_{\mathbf{A}}Q\left(\mathbf{S},\mathbf{A}|\theta^{Q}\right)|_{\mathbf{S}=\mathbf{S}(t),\mathbf{A}=\mu\left(\mathbf{S}(t)\right)}\nonumber\\
&\:\:\nabla_{\theta_{\mu}}\mu\left(\mathbf{S}|\theta^{\mu}\right)|_{\mathbf{S}=\mathbf{S}(t)}].
\end{align}

The loss function of the Q network is
\begin{align}\label{Q_loss}
\theta^{Q}=E_{\mu^{'}}\left[\left(Q\left(\mathbf{S}(t),\mathbf{A}(t)|\theta^{Q}\right)-y(t)\right)^{2}\right],
\end{align}
where
\begin{align}\label{yi}
\resizebox{.87\hsize}{!}{$y(t)=r\left(\mathbf{S}(t),\mathbf{A}(t)\right)+\gamma Q\left(\mathbf{S}(t+1),\mu\left(\mathbf{S}(t+1)\right)|\theta^{Q}\right)$}.
\end{align}

\subsection{Proposed P-DDPG Algorithm}
In RL, every action taken by the agent needs to consider the expected reward of the action in the future. However, the state/action spaces are very large in the actual high-mobility communication systems. As a result, the agent is difficult to explore and converge in the limited time. Therefore, we propose a novel P-DDPG algorithm, which can reduce the exploratory number of agent and accelerate the convergence speed by providing more prior information to the agent.

The traditional Q value is updated via (\ref{Q_update}), where $r(\mathbf{S}(t),\mathbf{A}(t))$ denotes the immediate reward value of action $\mathbf{A}(t)$ in state $\mathbf{S}(t)$, and $Q^{\mu}\left(\mathbf{S}(t+1),\mu\left(\mathbf{S}(t+1)\right)\right)$ is the reward of actions $\mu\left(\mathbf{S}(t+1)\right)$ taken in state $\mathbf{S}(t+1)$ after action $\mathbf{A}(t)$ is performed. In our proposed P-DDPG algorithm, (\ref{Q_update}) is redefined as
\begin{align}\label{newQ_update}
Q^{\mu}\left(\mathbf{S}(t),\mathbf{A}(t)\right)&=\mathbb{E}_{r_{t},\mathbf{S}(t+1)\sim E}[r(\mathbf{S}(t),\mathbf{A}(t))\nonumber\\
&+(1-\varrho(t))\gamma Q^{\mu}\left(\mathbf{S}(t+1),\mu\left(\mathbf{S}(t+1)\right)\right)\nonumber\\
&+\varrho(t) Q^{\mu}_{\mathrm{pre}}\left(\mathbf{S}_{\mathrm{pre}}(t+1),\mathbf{A}(t+1)\right)],
\end{align}
where $Q^{\mu}_{\mathrm{pre}}\left(\mathbf{S}_{\mathrm{pre}}(t\!+\!1),\mathbf{A}(t\!+\!1)\right)$ denotes the future reward based on the prediction state $\mathbf{S}_{\mathrm{pre}}(t\!+\!1)$; and $\varrho(t)$ is confidence coefficient in the time-slot $t$, related to the prediction accuracy.

\begin{remark}
$\gamma$ actually has specific physical meanings. Firstly, since the considered system has the prediction characteristic, when evaluating a state, we not only calculate the effects of the current state, but also estimate the effects of the adjacent states. Secondly, $Q^{\mu}\left(\mathbf{S}(t+1),\mu\left(\mathbf{S}(t+1)\right)\right)$ represents the average reward inspired by the past historical memory. Therefore, the P-DDPG algorithm is a forward-looking method. It considers the past, the present and the future information, which can avoid falling into a local optimal solution.
\end{remark}

By applying proposed P-DDPG algorithm, a novel learning-based DMCA framework is elaborated in next section.
\section{Implementation of the Learning-based DMCA}
\begin{figure}[h]
	\centering{}\includegraphics[scale=0.65]{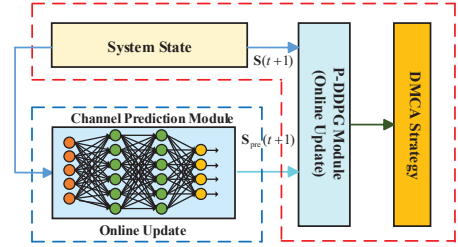}
	\caption{The learning-based DMCA overall framework.}\label{Dynamic}
\end{figure}

We know that P-DDPG is a new algorithm combining the prediction with DDPG. Therefore, the learning-based DMCA framework should consist of two modules, namely, the CPM module for the channel prediction and the P-DDPG module (leverages the P-DDPG algorithm to learn and update) for outputting final channel access policies. The learning-based DMCA overall framework is shown in Fig. \ref{Dynamic}. The recurrent neural network (RNN)-based CPM module is in the blue dashed box, and the P-DDPG module with input and output is in the red dashed box. In this section, we will introduce the specific implementation of the CPM and P-DDPG modules in detail.
\subsection{Channel Prediction Module}
For general-purpose sequence modeling, long-short term memory (LSTM) model, as a special RNN structure, has been proven to be stable and powerful for modeling long-range dependencies in various previous studies \cite{gers1999learning}. LSTM uses the memory cell and gates to control information flow, and thus, the gradient will be trapped in the cell (also known as constant error carousels) and prevented from vanishing too quickly, which is a critical problem for the vanilla RNN model. Based on LSTM, this paper designs an online CPM by combining with the incremental learning (IL) technology \cite{hong2008unsupervised,polikar2001learn++}.
\subsubsection{Channel Prediction Module Structure}
\begin{figure}[h]
	\centering{}\includegraphics[scale=0.5]{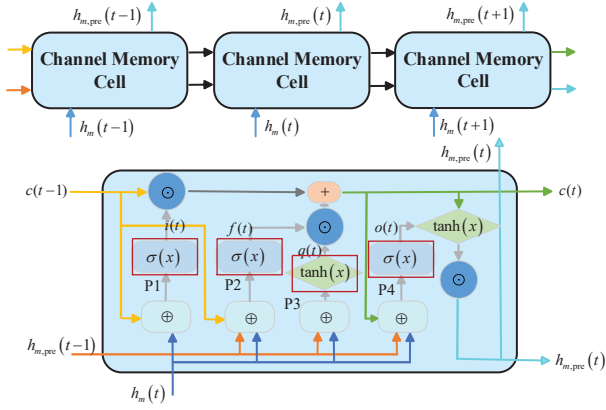}
	\caption{The structure of CPM. $\sigma(x)$ and $\tanh(x)$ are the activation functions, where $\sigma(x)=1/(1+e^{-x})$.}\label{LSTM_stru}
\end{figure}

According to the thought of LSTM, the structure of CPM is designed in Fig. \ref{LSTM_stru}, where the basic channel memory cell (i.e., the basic LSTM cell) is shown for clarification. Denote $\odot$ as the Hadamard product, and $\bigoplus$ represents matrix weighted sum.

For the basic channel memory cell, at every time a new input comes, its information will be accumulated to $c(t)$ if the input gate $i(t)$ is activated. Also, the past cell status $c(t)$ could be forgotten in this process if the forget gate $f(t)$ is on. Whether the latest cell output $c(t)$ will be propagated to the final state, $h(t)$ is further controlled by the output gate $o(t)$. Please refer to \cite{hochreiter1997long} for the detailed operations performed on the basic channel memory cell (the basic LSTM cell). Through the above operations, the channel memory cell can retain useful information and improve the prediction accuracy.

In each time-slot, the channel gain data $h_{m}(t)$ is a one-dimensional real number. Therefore, the specific structure of CPM can be described as follows. First, our CPM consists of three parts, i.e., input layer, middle LSTM layer and output layer. The input layer is a fully connected neural network (NN) only containing two layers. The number of neurons in the first layer is set to 1, and the number of neurons in the second layer is ``Unit number''\footnote{``Unit number'' refers to the number of neurons in the basic LSTM cell, and is also the number of output neurons. Specifically, in Fig. 3, operations P1, P2, P3 and P4 are all small feed-forward NNs. The number of neurons in the four feed-forward NNs is the ``Unit number''.}. For the middle LSTM layer, the basic cell structure is shown in Fig. \ref{LSTM_stru}. In this paper, the ``Time step''\footnote{``Time step'', i.e., parameter $k$, refers to the LSTM network that considers each input data to be related to the amount of the previous channel data. It is not difficult to see that ``Time step'' has similar meaning to parameter $k$ in the $k$-order Markov process, and hence ``Time step'' is represented by parameter $k$ in this paper.} and ``Unit number'' are both 5. The output layer is also a two-layer fully connected NN. The number of neurons in the first layer is ``Unit number'', and the number of neurons in the second layer is set to 1. Note that the input of the output layer is the channel memory cell state in the last ``Time step''.
\subsubsection{IL-Based Training Process}
In the general off-line prediction algorithm, the training process of the prediction model is performed as follows. First, the model is trained on the training set, and the validation set is used to adjust hyper-parameters. Then, the trained prediction model is tested on the test set. Different from the above learning process, our CPM is an online algorithm that alternately performs the training and prediction processes, thus it can effectively counter the fast time-varying channels.

Specifically, this paper considers two CPMs: single-point CPM (SPCPM) and multi-point CPM (MPCPM). SPCPM only predicts the channel data of the next one time-slot, while in MPCPM, the channel data of future multiple time-slots are predicted. Let $l$ denote the prediction length of CPM. $l=1$ means SPCPM, and $l>1$ refers to MPCPM.
\begin{itemize}
  \item \textbf{Implementation Process of SPCPM}:
\end{itemize}
\begin{figure}[h]
	\centering{}\includegraphics[scale=1.1]{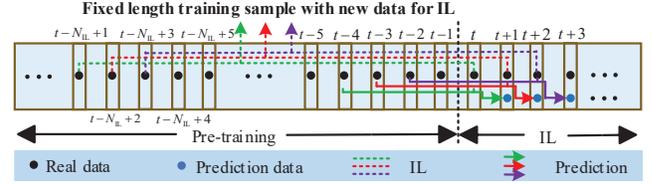}
	\caption{The implementation process of SPCPM.}\label{single_point_structure}
\end{figure}

Fig. \ref{single_point_structure} shows the implementation process of SPCPM, including the pre-training and IL phases. As shown in Fig. \ref{single_point_structure}, the implementation process can be described as follows:
\begin{enumerate}
      \item [i)] Leverage the historical data (from the 0-th to the ($t\!-\!1$)-th time-slots) to train, and acquire a pre-training model.
      \item [ii)] Update the model by IL at the $t$-th time-slot: gather the channel data of the $t$-th time-slot, and construct a fixed length training sample. The training sample contains the latest $N_{\mathrm{IL}}$ channel data, i.e., the ($t\!-\!N_{\mathrm{IL}}\!+\!1$)-th to the $t$-th time-slots. Then, the training sample is employed to train and update the parameters of pre-training model.
      \item [iii)] Predict the channel data of the ($t+1$)-th time-slot at the $t$-th time-slot: the channel data of the ($t-k+1$)-th to the $t$-th time-slots are input to the trained model in ii), and the channel data of the ($t+1$)-th time-slot is output.
      \item [iv)] Repeat ii) and iii) to continuously alternate the training and prediction processes in order.
\end{enumerate}
\begin{itemize}
  \item \textbf{Implementation Process of MPCPM}:
\end{itemize}

Similar to SPCPM, the implementation process of MPCPM is described as follows:
\begin{enumerate}
      \item [i)] Leverage the historical data (from the 0-th to the ($t\!-\!1$)-th time-slots) to train, and acquire a pre-training model.
      \item [ii)] Update model by IL at the $t$-th time-slot: gather the channel data of the $t$-th time-slot, and update the parameters of the pre-training model based on the training sample containing the latest $N_{\mathrm{IL}}$ data by the IL method.
      \item [iii)] Predict the channel data of the ($t+1$)-th to the ($t+l$)-th time-slots at the $t$-th time-slot: first, the channel data of the ($t-k+1$)-th to the $t$-th time-slots are input to the trained model in ii), and the channel data of the ($t+1$)-th time-slot is the output. Then, the channel data of the ($t-k+2$)-th to the ($t+1$)-th time-slots are input to the same model and the channel data of the ($t+2$)-th time-slot is the output, where the used data of the ($t+1$)-th time-slot is the value that has been predicted. Next, the above flow is repeated until the ($t+1$) to the ($t+l$) time-slots data are all predicted.
      \item [iv)] Repeat ii) and iii) to continuously alternate the training and prediction processes in order. It should be noted that the next process of updating the model by IL occurs in the ($t+l$)-th time-slot.
\end{enumerate}
\subsection{P-DDPG Module}
The implementation of the P-DDPG module consists of three parts, i.e., state space, action space, and reward function.
\subsubsection{Design of Action Space}
In the P-DDPG policy network, a user action is represented by a set of neurons. Therefore, two major difficulties are described based on the characteristics of DMCA:
\begin{itemize}
  \item Difficulty 1: As users ate moving, the number of users $N$ changes, leading to the changes of the network structure. However, the network needs to be retained when the network structure changes, which is impossible in practice, due to the training cost.
  \item Difficulty 2: The one-hot code is commonly used in the action representation. Hence, $M$ channels imply that there are $M$ choices for each user, and then $M$ neurons are needed to describe each action. So, in the system with $N$ users, $MN$ neurons are needed, and this number becomes larger as $M$ and $N$ increases.
\end{itemize}

For difficulty 1, the concept of \emph{virtual user} is proposed, which is a user whose requirement is $R_{\mathrm{user}}(t)=0$. Assume that the system can serve up to $K$ users at the same time ($K\leq M$ and $N\leq K$). When there are $N$ users in the system, there are $K-N$ users with $R_{\mathrm{user}}(t)=0$. In this way, there are always $K$ users in the system, ensuring the stability of the network, which indicates that the number of output neurons is always $MK$.

For difficulty 2, we only use a neuron to represent the $M$ action choices of a user. Specifically, the output values of the output neurons are first normalized to $\left(0,1\right)$, and then quantized to $M$ levels. The output value of each neuron is a continuous value between $(0,1)$ before quantification, which ensures that the action space is large enough to be fully explored.

Finally, only $K$ output neurons are needed after applying the \emph{virtual users} and $M$-level quantification.
\subsubsection{Design of State Space}
In the previous researches, to facilitate modeling and analysis, the channel model is usually assumed as the Markov model with two states or the $X$ states. Our learning-based DMCA framework does not require specific mathematical expressions, and thereby can be applied and analyzed directly even for continuous state spaces.

Although the system states vary in DMCA, the system states should always contain two parts: the channel service rate $\mathbf{S}_{\mathrm{ch}}(t)=\left[R_{1}(t),R_{2}(t),\cdots,R_{N}(t)\right]$ and user requirement rate $\mathbf{S}_{\mathrm{user}}(t)=\left[R_{\mathrm{user},1}(t),R_{\mathrm{user},2}(t),\cdots,R_{\mathrm{user},N}(t)\right]$. Therefore, the state is described as $\mathbf{S}(t)=[\mathbf{S}_{\mathrm{ch}}(t),\mathbf{S}_{\mathrm{user}}(t)]$.

However, the above $\mathbf{S}(t)$ cannot completely describe the state space because the prediction information is considered in P-DDPG. Thus, the state space should be $\mathbf{S}(t)=[\mathbf{S}_{\mathrm{ch}}(t),\mathbf{S}_{\mathrm{user}}(t),\mathbf{S}_{\mathrm{pre}}(t+1),\cdots,\mathbf{S}_{\mathrm{pre}}(t+l)]$, where $\mathbf{\mathbf{S}_{\mathrm{pre}}}(t+l)=\left[R_{\mathrm{pre},1}(t+l),R_{\mathrm{pre},2}(t+l),\cdots,R_{\mathrm{pre},M}(t+l)\right]$.

\subsubsection{Paradigm of Reward Function Design}
In existing studies, a few simple discrete reward values are adopted. For example, the successful event is denoted as 1, while the unsuccessful one is 0. Inspired by the process of the human learning, if the teacher can give more careful guidance and more prompt correction for us, we can learn faster. The \emph{careful guidance and prompt correction} corresponding to the RL is the reward function. Therefore, continuous and effective reward values can avoid the disadvantages of spare reward in traditional methods, further accelerating the convergence rate.

\begin{itemize}
  \item \textbf{Reward of LSM}:
\end{itemize}
\begin{enumerate}
  \item [i)] $\vartheta(t)=1$, i.e., $R_{\mathcal{A}_{n}(t)}(t)\geq R_{\mathrm{user},n}(t)$ for $\forall n$. Then, $r(t)$ can be defined as
\begin{align}
r(t)=\varpi_{1}\mathrm{arccot}(\varpi_{2}(\underset{n\in\mathcal{N}}{\sum}(\arctan(\varDelta_{n}(t)))),
\end{align}
where $\varpi_{1}$ and $\varpi_{2}$ are the weight factors. $\arctan(x)$ is used to weaken the bias caused by some too large values of $R_{\mathcal{A}_{n}(t)}(t)-R_{\mathrm{user},n}(t)$.
$\mathrm{arccot}(x)$ is to amplify the small error, and thus ensues the validity of the reward.
  \item [ii)] $\vartheta(t)\in[0,1)$, i.e., $R_{\mathcal{A}_{n}(t)}(t)<R_{\mathrm{user},n}(t)$ for $\exists n$. Then, $r(t)$ can be expressed as
\begin{align}
r(t)=\vartheta(t)\varpi_{1}\mathrm{arccot}(\varpi_{2}\varSigma),
\end{align}
where
\begin{align}
&\varSigma=\underset{n\in\mathcal{N}}{\sum}(\arctan(|\varDelta_{n}(t)|/\varUpsilon(\varDelta_{n}(t))),\nonumber\\
&\varUpsilon(\varDelta_{n}(t))=\begin{cases}
1 & \varDelta_{n}(t)\geq0;\\
\vartheta(t)+\varepsilon & \varDelta_{n}(t)<0,
\end{cases}
\end{align}
and $\varepsilon$ is very small, such as $10^{-7}$, preventing the denominator from being 0.
\end{enumerate}

In addition, the reward function needs to prevent the policy network from allocating the same channels to different users. Therefore, the final reward function is given by
\begin{align}\label{living_reward}
r_{\mathrm{final}}(t)=\mathcal{H}(\mathbf{A}(t))r(t)+(1-\mathcal{H}(\mathbf{A}(t)))\varpi_{3},
\end{align}
where
\begin{align}
\mathcal{H}(\mathbf{A}(t))=\begin{cases}
1 & \mathcal{A}_{i}(t)\neq\mathcal{A}_{j}(t)\:\mathrm{for}\:\forall i,j,i\neq j;\\
0 & \mathcal{A}_{i}(t)\mathcal{=A}_{j}(t)\:\mathrm{for}\:\exists i,j,i\neq j,
\end{cases}
\end{align}
and $\varpi_{3}$ is a penalty factor, and $\varpi_{1},\varpi_{2}>0$, $\varpi_{3}<0$.

\begin{itemize}
  \item \textbf{Reward of BSM}:
\end{itemize}
\begin{enumerate}
  \item [i)] $\vartheta(t)=1$, $r(t)$ can be described as
\begin{align}
r(t)=\varpi_{1}(\exp(\alpha_{1}\varpi_{2}\varLambda)-1),
\end{align}
where
\begin{align}
\varLambda=\underset{n\in\mathcal{N}}{\sum}(\arctan(\varDelta_{n}(t)),
\end{align}
and $\alpha_{1}$ is a adjustable factor.
  \item [ii)] $\vartheta(t)\in[0,1)$, and $r(t)$ can be defined as
\begin{align}
r(t)=\vartheta(t)\varpi_{1}(\exp(\alpha_{1}\varpi_{2}\varGamma)-1),
\end{align}
where
\begin{align}
\varGamma=\exp(\alpha_{2}\underset{n\in\mathcal{N}}{\sum}(\varDelta_{n}(t)/\varUpsilon(\varDelta_{n}(t)))),
\end{align}
and $\alpha_{2}$ is also a adjustable factor, satisfying $\alpha_{1},\alpha_{2}>0$.
\end{enumerate}

The expression of the final reward function in BSM is consistent with (\ref{living_reward}), which will not be specified here.

\begin{remark}
It is emphasized that when RL is used to solve the optimization problem in wireless communications, the objective function can be transformed into the design of one part of the reward function, and the constraints are converted into the designs of the action space, the state space and another part of the reward function.
\end{remark}
\subsubsection{Training Process of the Learning-Based DMCA Method}
The proposed learning-based DMCA scheme is summarized in algorithm \ref{alg:Framwork}.
\begin{algorithm}[h]
\caption{The learning-based DMCA method}.
\label{alg:Framwork}
\begin{algorithmic}[1] 
\REQUIRE ~~\\ 
Randomly initialize the critic network $\theta^{Q}$/$\theta^{Q^{'}}$, and the actor network $\theta^{\mu}$/$\theta^{\mu^{'}}$ for the online/target network\\
Initialize experience replay pool $\mathcal{B}$
\ENSURE ~~\\ 
\FOR{episode = 1,...,$E$}
\STATE Initialize $\mathbf{A}(0)=\mathbf{0}$
\STATE Initialize a random process $\varPsi(t)$ for action exploration
\STATE Receive initial observation state $\mathbf{S}(t)$
\FOR{steps = 1,...,$L$}
\STATE Select action $\mathbf{A}(t)=\mu\left(\mathbf{S}(t)|\theta^{\mu}\right)+\varPsi(t)$
\STATE Execute action $\mathbf{A}(t)$, observe reward $r_{\mathrm{final}}(t)$, and observe new state $\mathbf{S}(t+1)$
\STATE Store transition $(\mathbf{S}(t),\mathbf{A}(t),r_{\mathrm{final}}(t),\mathbf{S}(t+1))$ in $\mathcal{B}$
\STATE Sample a random mini-batch of $I$ transitions $(\mathbf{S}(i),\mathbf{A}(i),r_{\mathrm{final}}(i),\mathbf{S}(i+1))$ from $\mathcal{B}$
\STATE Update critic by minimizing the loss by (\ref{Q_loss})
\STATE Update the actor using the sampled gradient by (\ref{Policy_loss})
\STATE Soft update the target networks
\STATE Satisfy the double stop criterion: \textbf{Break}
\ENDFOR
\ENDFOR
\end{algorithmic}
\end{algorithm}
\begin{remark}
``Soft update'' refers to the way of being updated step by step, and not completely replaced, that is, $\theta^{Q^{'}}\leftarrow\tau\theta^{Q}+\left(1-\tau\right)\theta^{Q^{'}}$ and $\theta^{\mu^{'}}\leftarrow\tau\theta^{\mu}+\left(1-\tau\right)\theta^{\mu^{'}}$, where $\tau\ll1$. The random process $\varPsi(t)$ usually adopts the Ornstein-Uhlenbeck process.
\end{remark}

We can see that the proposed learning-based DMCA framework is very complicated. For the actual deployment, the corresponding compression and acceleration technologies of neural networks are essential, and some effective technologies have been proposed in \cite{guo2019compression}. However, considering the scope of this paper, it is our future direction to employ effective compression acceleration technologies to promote the actual deployment of our framework.
\section{Simulation Results}
In this section, we verify the performance of the proposed learning-based DMCA scheme in various scenarios for the single- and multi-user access. In simulation, the noise power is $\sigma_{m}^{2}=-125$ dBm, and the each channel bandwidth is $B_{m}=78.125$ KHz. The total power is $P_{\mathrm{max}}\!=\!43$ dBm, which is evenly distributed to all channels. For channel parameter $h_{m}(t)$, the test date provided by \emph{Huawei} is directly applied.

In simulations, the used hyper-parameters are also given in table \uppercase\expandafter{\romannumeral1} in detail, unless otherwise specified.
\begin{table}[h]
	\centering{}
	\textbf{Table \uppercase\expandafter{\romannumeral1}}~~The hyper-parameters of the learning-based DMCA.\\
	\setlength{\tabcolsep}{0.9mm}{
		\begin{tabular}{cc|cc|cc} \toprule
			Parameter   & Value  & Parameter & Value & Parameter & Value\\
			\midrule
			$\varpi_{1}$/$\varpi_{2}$          & 5        & $I$   & 32      & $\varpi_{3}$  & -100\\
			$\alpha_{1}$               &  0.5   & $\alpha_{2}$   & 0.5   & $\beta$            & 0.8\\
            $\mathcal{B}$ Size & 2000  & $ l $                 & 5      &$\gamma$        & 0.92 \\
            Mini-batch size          & 64      & E                      & 300  &$L$                  & 3000\\
            \bottomrule
	\end{tabular}}
\end{table}

\begin{figure*}[htb]
  \subfigure[]{
    \includegraphics[width=0.5\textwidth, height=20mm]{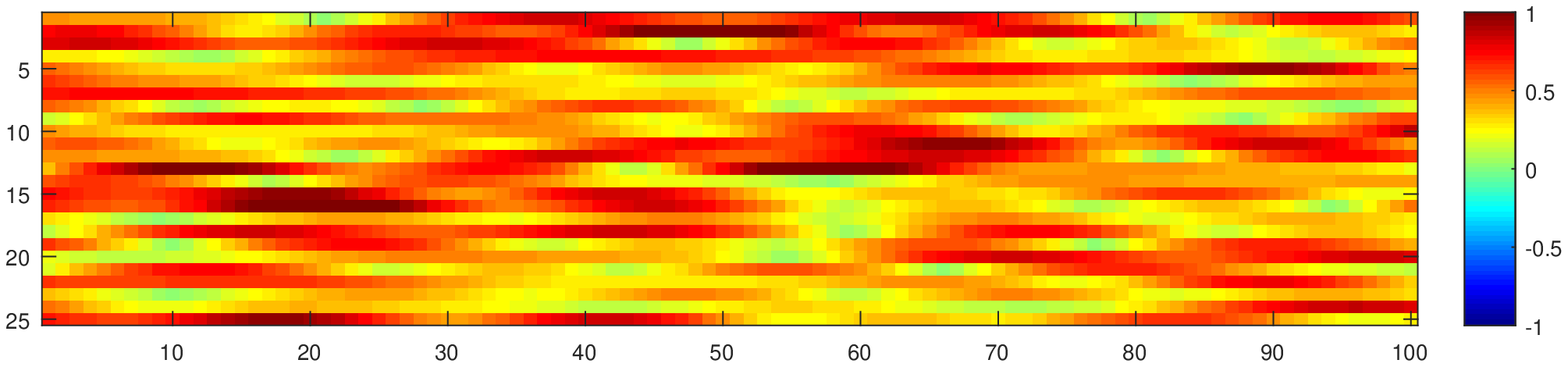}
  }
  \subfigure[]{
    \includegraphics[width=0.5\textwidth, height=20mm]{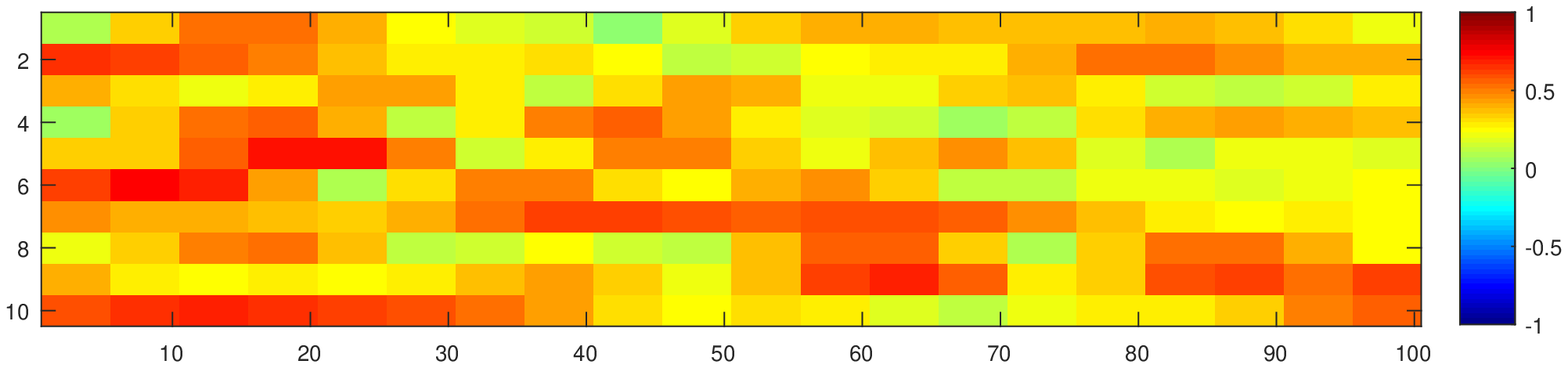}
  }
  \caption{(a) is the change of 25 channels $h_{m}(t)$ in 100 time-slots, and (b) is the change of the user requirement factor $\xi_{n}(t)$ of 10 users in 100 time-slots.}
  \label{channel_user_state} 
\end{figure*}

In order to facilitate presentation, 25 channels ($M=25$) are used for user selections, and the corresponding changes of channel gains $h_{m}(t)$ ($m\in\mathcal{M}$) are shown in Fig. \ref{channel_user_state}(a). Assume that the system can serve up to 10 users ($K\!=\!10$) simultaneously, and the changes of the user requirements is shown in Fig. \ref{channel_user_state}(b)\footnote{Fig. \ref{channel_user_state}(b) shows the changes in user requirement factor $\xi_{n}(t)$, and the actual user requirement is $R_{\mathrm{user},n}(t)=\rho B_{n}\log\left(1+\frac{|\xi_{n}(t)|^{2}P_{n}}{\sigma_{n}^{2}}\right)$, where $\rho$ is the weighted factor and $\rho=0.9$ in this paper.}. The \emph{delay sensitivity} $\lambda_{n}$ of the 10 users is $\{0.85,0.95,0.82,0.94,0.63,1.00,0.76,0.91,0.89,0.81\}$. It should be pointed out that unlike the channels changing every time-slot, the user requirements remain unchanged for a longer time period, and the duration is 5 time-slots in this paper.

Next, we successively verify the prediction performance of CPM, the effectiveness of the DMCA framework in single- and multi-user scenarios, the stability of the service, and the effects of prediction on the convergence rate.
\subsection{Performance comparison of CPM}
In this subsection, we verify the prediction performance of our proposed SPCPM and MPCPM by applying the real channel data, and we also explore the impact of different ``Time step'' $k$ and prediction length $l$ on the prediction results. To provide subsequent researchers with a baseline, the corresponding code and data have been posted in \cite{github}\footnote{The data set is composed of four sub-sets. In the first sub-set, the motion speed is $v=180$ km/h, the carrier frequency is $f_{c}=3$ GHz, and the channel sampling rate is $f_{s}=200$ kHz. The channel data of the remaining sub-sets are $v=90$ km/h, $v=360$ km/h, and $v=450$ km/h, respectively, while other parameters remain unchanged. At the same time, each sub-set contains fading changes of $9$ multi-path channels.}.

For channels at different speeds, the channel data within 40 ms are considered. Since the channel sampling rate is 200 kHz, there are a total of 8000 points. Among them, the first 6400 points are used for the pre-training phase, and the remaining 1600 points are used for the IL phase and testing. For the 6400 points in the pre-training phase, 4800 points are used in the training set, and the remaining 1600 points are used in the validation set. Meanwhile, the used hyper-parameters are given in Table \uppercase\expandafter{\romannumeral2}, unless otherwise specified.
\begin{table}[h]
	\centering{}
	\textbf{Table \uppercase\expandafter{\romannumeral2}}~~The hyper-parameters of CPM.
	\setlength{\tabcolsep}{1.1mm}{
		\begin{tabular}{cc|cc} \toprule
			Parameter   & Value  & Parameter & Value\\
			\midrule
			Learning rate   & 0.06        & $N_{\mathrm{IL}}$   & 200\\
			Input size    &  1   & Output size   & 1 \\
            Time step & 5 & Unit number & 5\\
            $l$ (SPCPM) & 1   & Iteration step (Pre-training) & 200 \\
            $l$ (MPCPM)   & 5 & Iteration step ( once IL) & 1\\
            \bottomrule
	\end{tabular}}
\end{table}
\begin{figure*}[htb]
	\centering{}\includegraphics[scale=0.4]{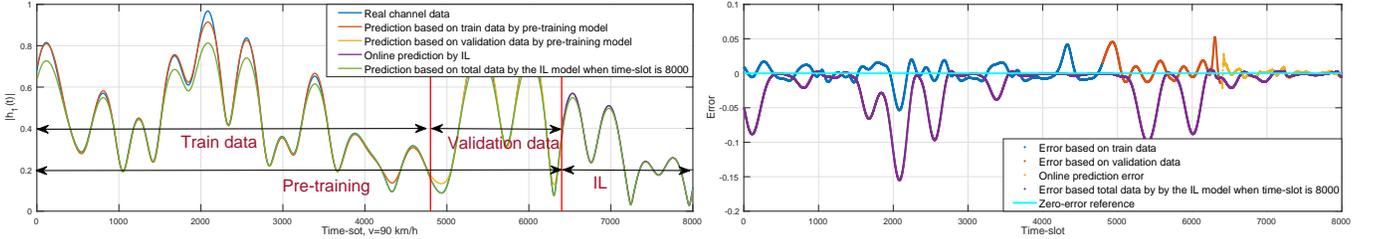}
	\caption{The prediction results of SPCPM for the 90 km/h channel.}\label{single_90_prediction}
\end{figure*}

\begin{figure}[htb]
	\centering{}\includegraphics[scale=0.5]{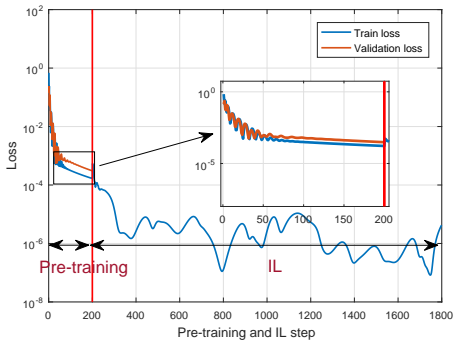}
	\caption{The loss curves of SPCPM for the 90 km/h channel. Note that the logarithmic coordinate system is used for a clearer display considering that the train loss drops very quickly.}\label{single_90_loss}
\end{figure}

Fig. \ref{single_90_prediction} depicts the prediction results of the pre-training phase and the IL phase for SPCPM. In order to observe and compare prediction results more clearly, the prediction error of each time-slot is also given. It can be seen that SPCPM has low prediction error. Note that SPCPM is an online prediction model based on IL, that is, the model parameters will change as time goes by. Therefore, we employ the trained model in the 8000-th time-slot to predict all previous channel data. The farther the channel data is from the current time-slot, the larger the prediction error is. It reveals the fact that SPCPM has strong real-time performance and can cope with the changing characteristics of fast time-varying channels. In addition, the loss changes are shown in Fig. \ref{single_90_loss}, and we can see that SPCPM has a good convergence performance.

\begin{figure*}[htb]
	\centering{}\includegraphics[scale=0.4]{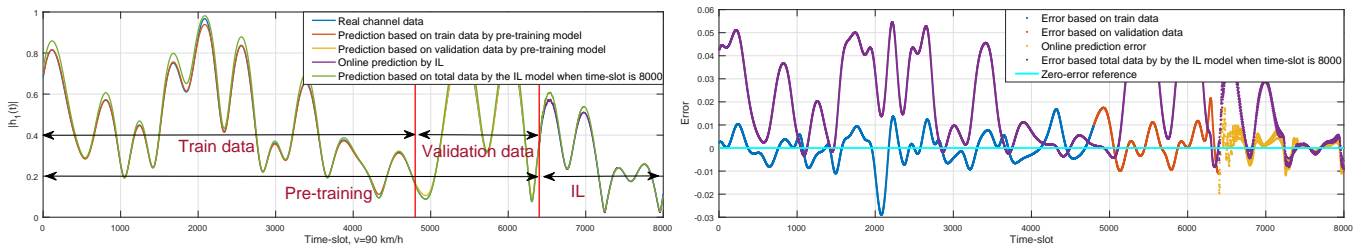}
	\caption{The prediction results of MPCPM for the 90 km/h channel.}\label{multi_90_prediction}
\end{figure*}

\begin{figure}[htb]
	\centering{}\includegraphics[scale=0.5]{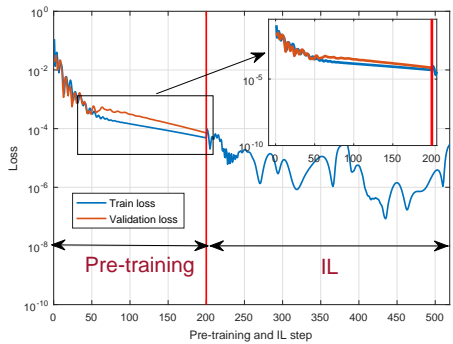}
	\caption{The loss curves of MPCPM for the 90 km/h channel.}\label{multi_90_loss}
\end{figure}

Meanwhile, MPCPM also show low prediction error and the fast convergence performance, as shown in Fig. \ref{multi_90_prediction} and Fig. \ref{multi_90_loss}.

\begin{figure*}[htb]
	\centering{}\includegraphics[scale=0.4]{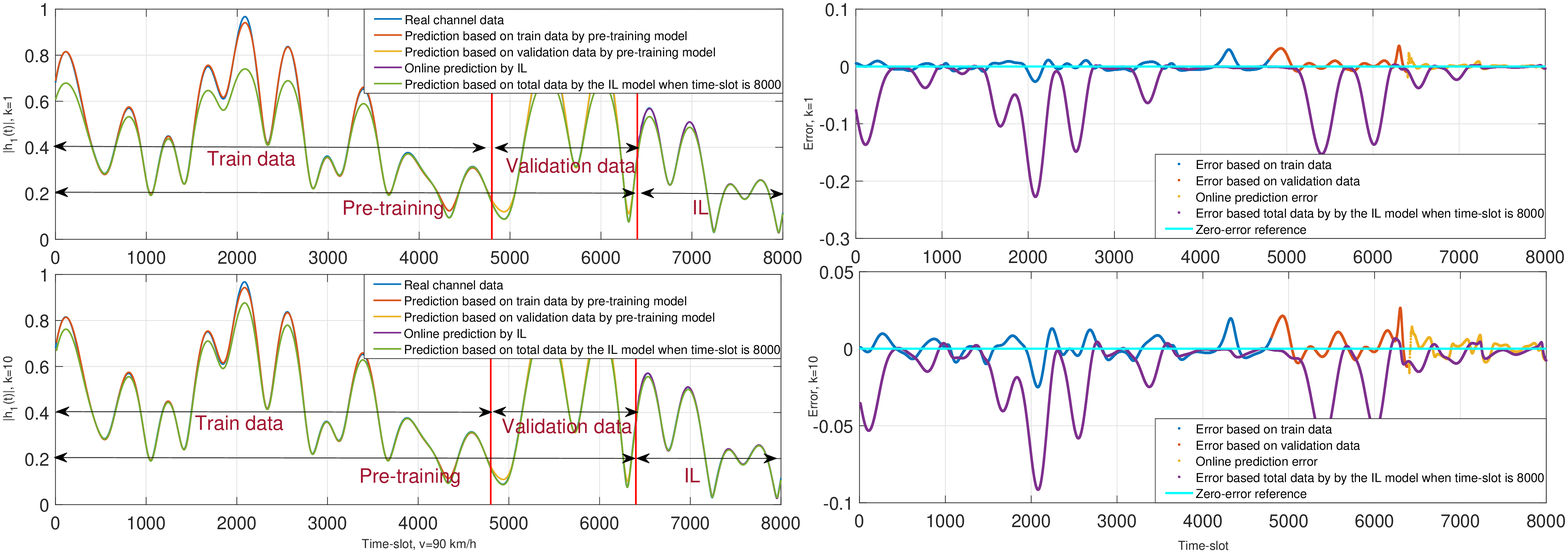}
	\caption{The prediction results of SPCPM for the 90 km/h channel under $k=1$ and $k=10$.}\label{single_90_prediction_time_step}
\end{figure*}

\begin{figure*}[htb]
 \centering{}
  \subfigure[]{
    \includegraphics[scale=0.5]{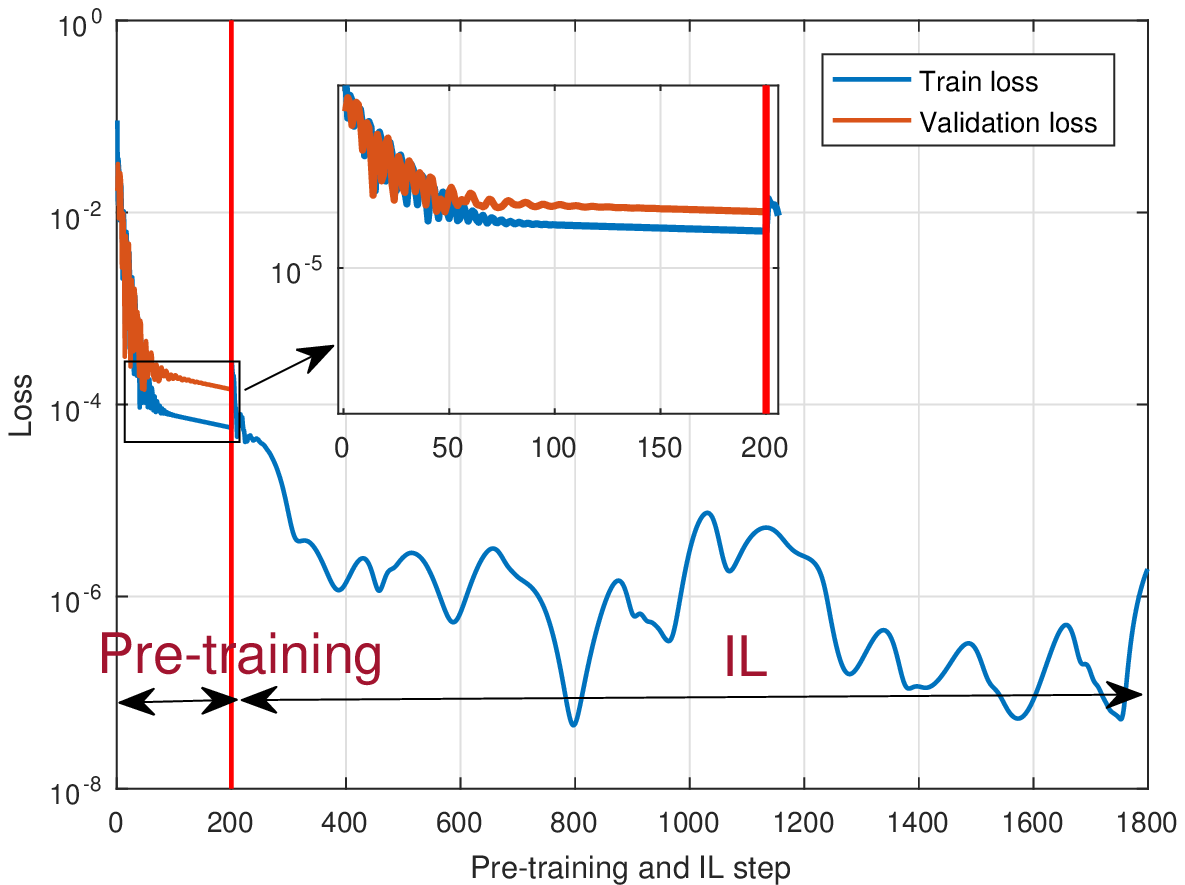}
  }
  \subfigure[]{
    \includegraphics[scale=0.5]{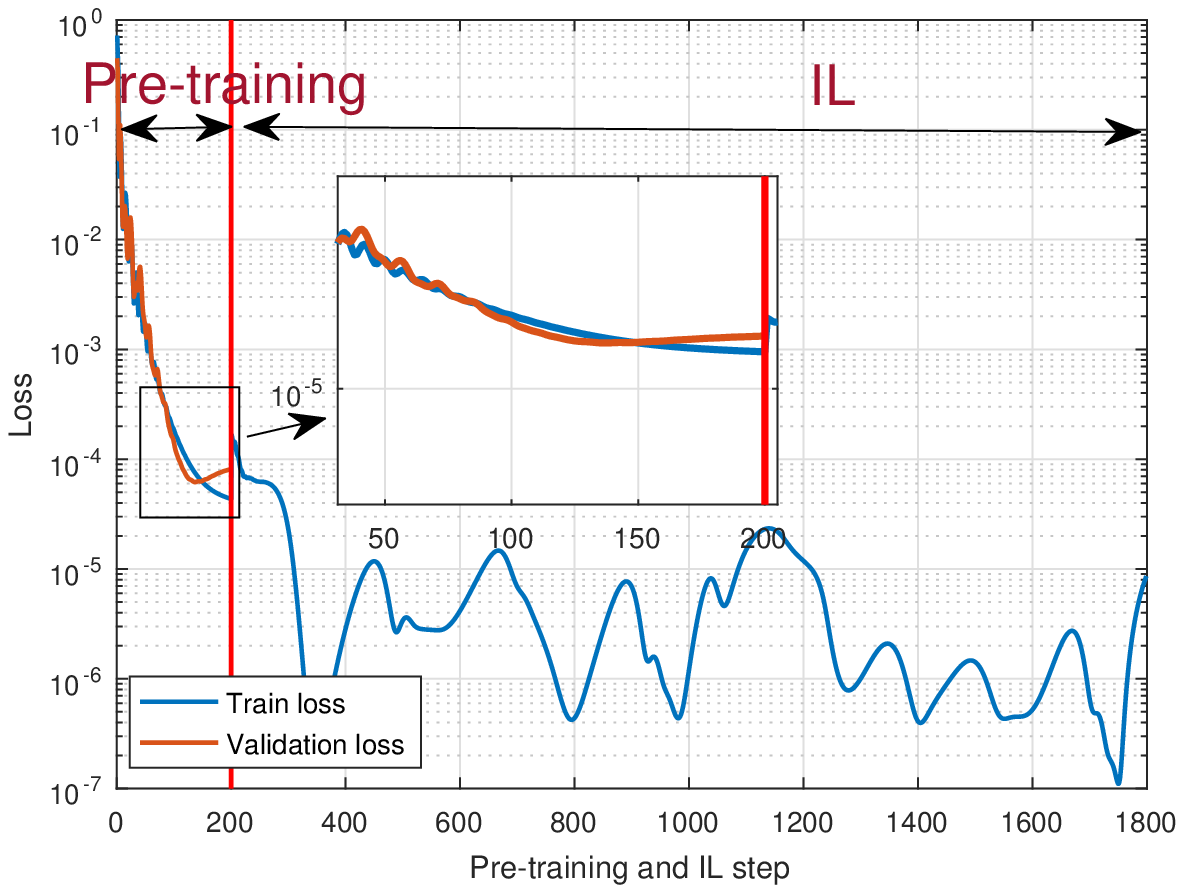}
  }
  \caption{The loss curves of SPCPM for the 90 km/h channel under $k=1$ (a) and $k=10$ (b).}
  \label{single_90_loss_time_step} 
\end{figure*}

Next, we will discuss the impact of different ``Time step'' on prediction results. Fig. \ref{single_90_prediction_time_step} and Fig. \ref{single_90_loss_time_step} show the prediction results and loss curves for SPCPM versus the 90 km/h channel when ``Time step'' is 1 or 10.

From Fig. \ref{single_90_prediction} and Fig. \ref{single_90_prediction_time_step}, we can see that our method owns good prediction performance under different $k$. As $k$ increases, the prediction error becomes smaller. This is mainly because with larger $k$, it is easier to find the correlation of the channel data over the time horizon. Nevertheless, considering the fact that the training of the model becomes slow as $k$ increases, larger $k$ is not reasonable. For instance, assume that $k$ is 5 in practice. If $k$ is larger than 5 in the prediction model, it will fail to improve the prediction accuracy, but can increase the training time of the model. Therefore, a reasonable value of $k$ is necessary for the performance of the model, including the prediction error and training speed.
\begin{figure*}[htb]
	\centering{}\includegraphics[scale=0.4]{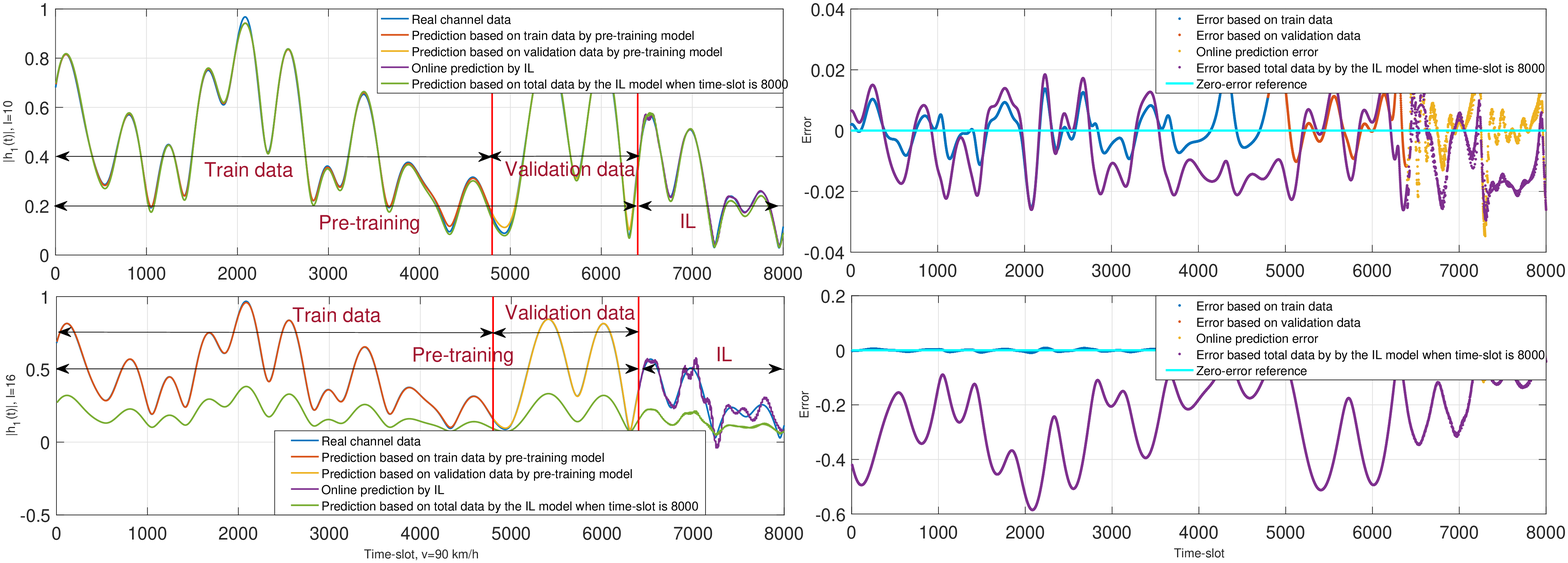}
	\caption{The prediction results of MPCPM for the 90 km/h channel under $l=10$ and $l=16$.}\label{multi_90_prediction_length}
\end{figure*}
\begin{figure*}[htb]
 \centering{}
  \subfigure[]{
    \includegraphics[scale=0.5]{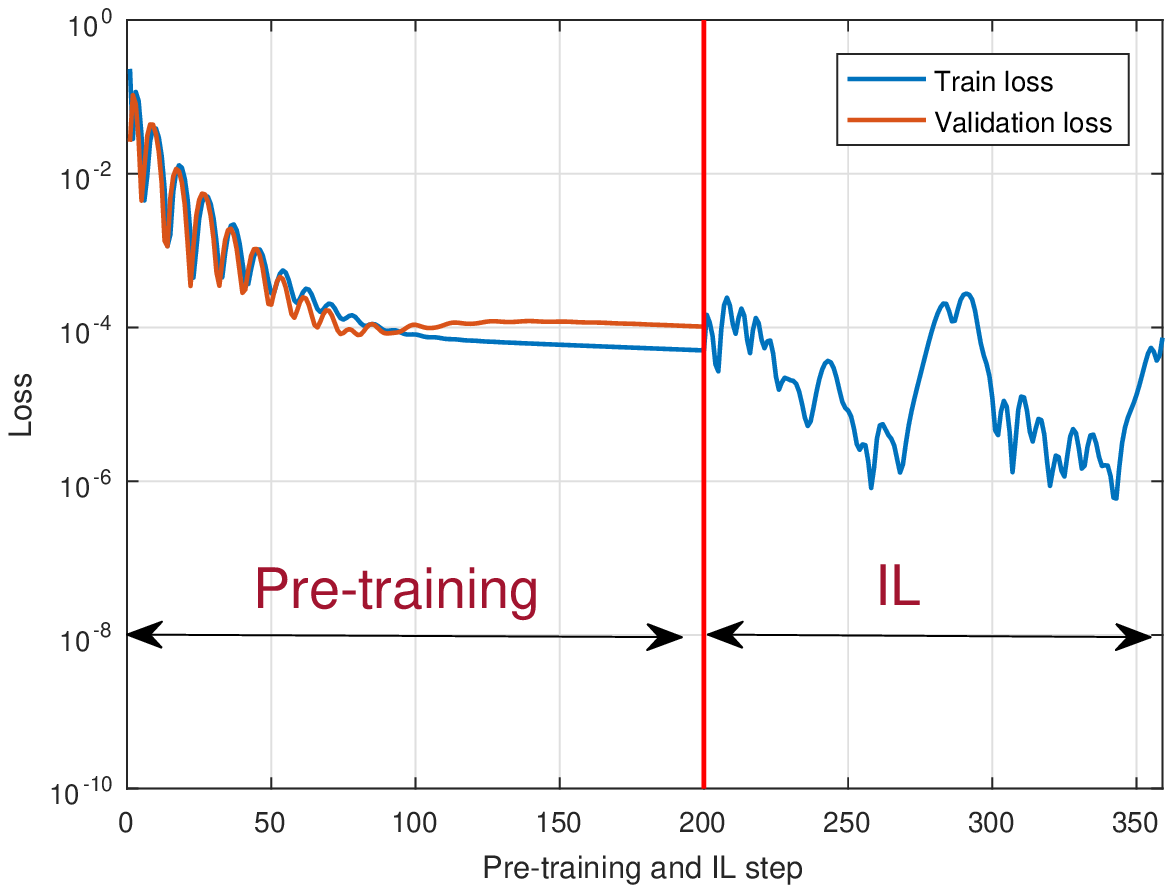}
  }
  \subfigure[]{
    \includegraphics[scale=0.5]{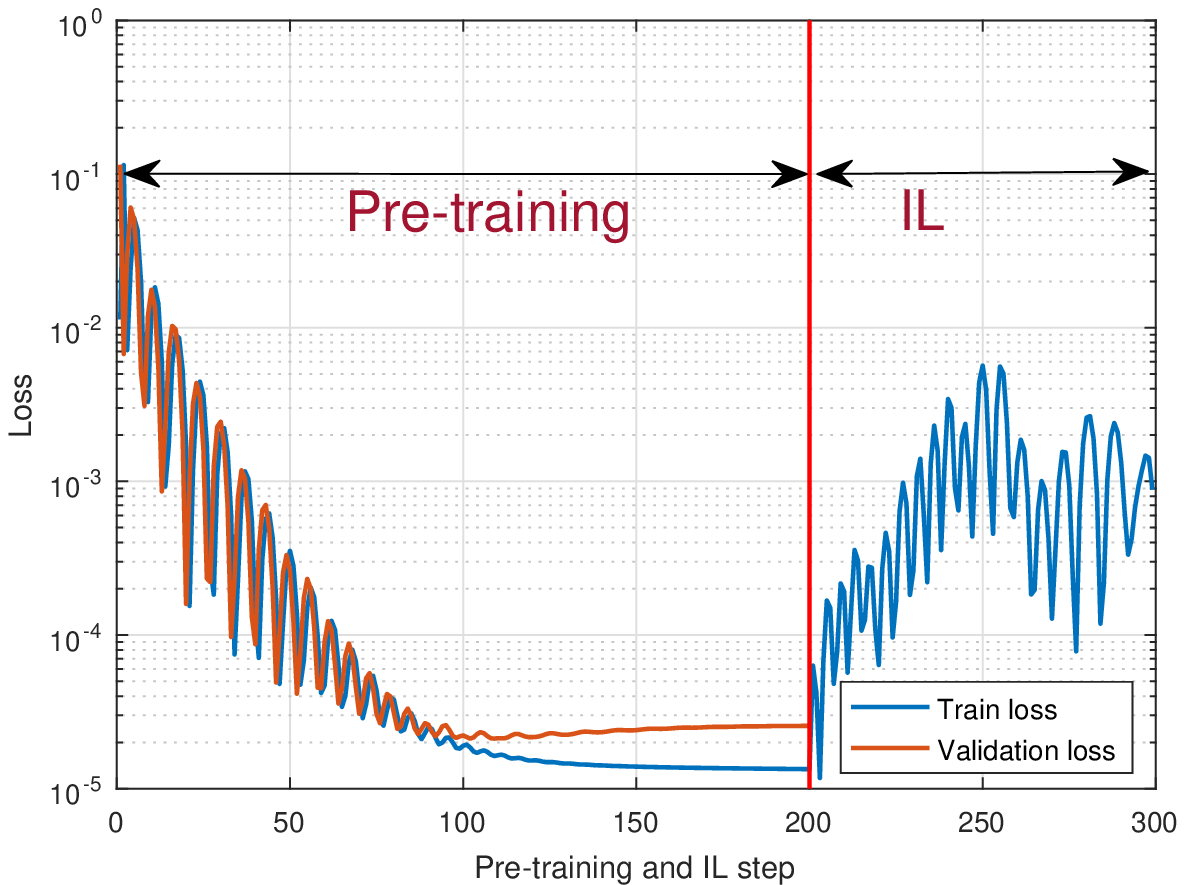}
  }
  \caption{The loss curves of MPCPM for the 90 km/h channel under $l=1$ (a) and $l=16$ (b).}
  \label{multi_90_loss_length} 
\end{figure*}

Fig. \ref{multi_90_prediction_length} and Fig. \ref{multi_90_loss_length} show the prediction and loss curves for different prediction lengths ($l=10$ and $l=16$). It can be seen that longer prediction length leads to larger error, which is in line with our expectations.

The experimental results with low errors, in addition to benefiting form our well-designed online CPM, also show that channel changes are predictable in a very short term. In our view, first, the channel changes are closely dependent on the environmental factors, including the frequency band, location, time, temperature, humidity, weather and so on \cite{8395053}. These parameters change slowly in a short period of time \cite{718494}. In addition, the channel data used in this paper is based on the vehicle environment with the speeds of 90 km/h, 180 km/h and 360 km/h. Considering that the vehicle usually travels in a fixed direction for a considerable period of time, the sport mode of the vehicle is hence relatively stable \cite{7867848}. Meanwhile, note that we only predict the magnitude of the channel gain, not the phase and other parameters. Based on the above, we believe that the channel variation between adjacent time-slots has a certain correlation in the short term. Considering that the parameters of CPM are constantly changing through IL, it can be explained that the channel change in a short term can be well predicted.

When modeling fast time-varying channels, compared to the \emph{Basis Expansion Model} \cite{rabbi2010high} and the first-order Taylor expansion-based predictive channel
modeling \cite{8093592}, although our CPM does not have detailed mathematical expressions, the performance is extremely excellent, which can be used to guide channel estimation/modeling. For example, our method can be easily combined with the channel estimation method proposed in \cite{8778685}, which may further improve the performance of channel estimation in corresponding scenarios.
\subsection{Single-User Access}
We have compared the learning-based scheme with the random\footnote{The random method is introduced in Section \uppercase\expandafter{\romannumeral1}.} and exhaustive search methods under \emph{one time-slot one decision} and \emph{$l$ ($l>1$) time-slots one decision}\footnote{\emph{One time-slot one decision}: An access policy is selected in each time-slot $t$. \emph{$l$ time-slots one decision}: An access policy is produced in time-slot $t$, and the next access policy is produced in time-slot $t+l$. The access policy remains unchanged from time-slot $t$ to time-slot $t+l$.} principles. $l=5$ in the simulations of this paper. The exhaustive search method means that $M!/(M-K)!$ cases are explored, and then the optimal access policy is selected at every time-slot $t$, which is used as a benchmark.
\subsubsection{Single-User Access in  LSM}
In LSM, the policy that satisfies criterion 1 is exactly what we need. Meanwhile, (\ref{step1}) indicates that $\vartheta(t)=1$ is the foremost goal that should be satisfied.
\begin{figure}[h]
	\centering{}\includegraphics[scale=0.55]{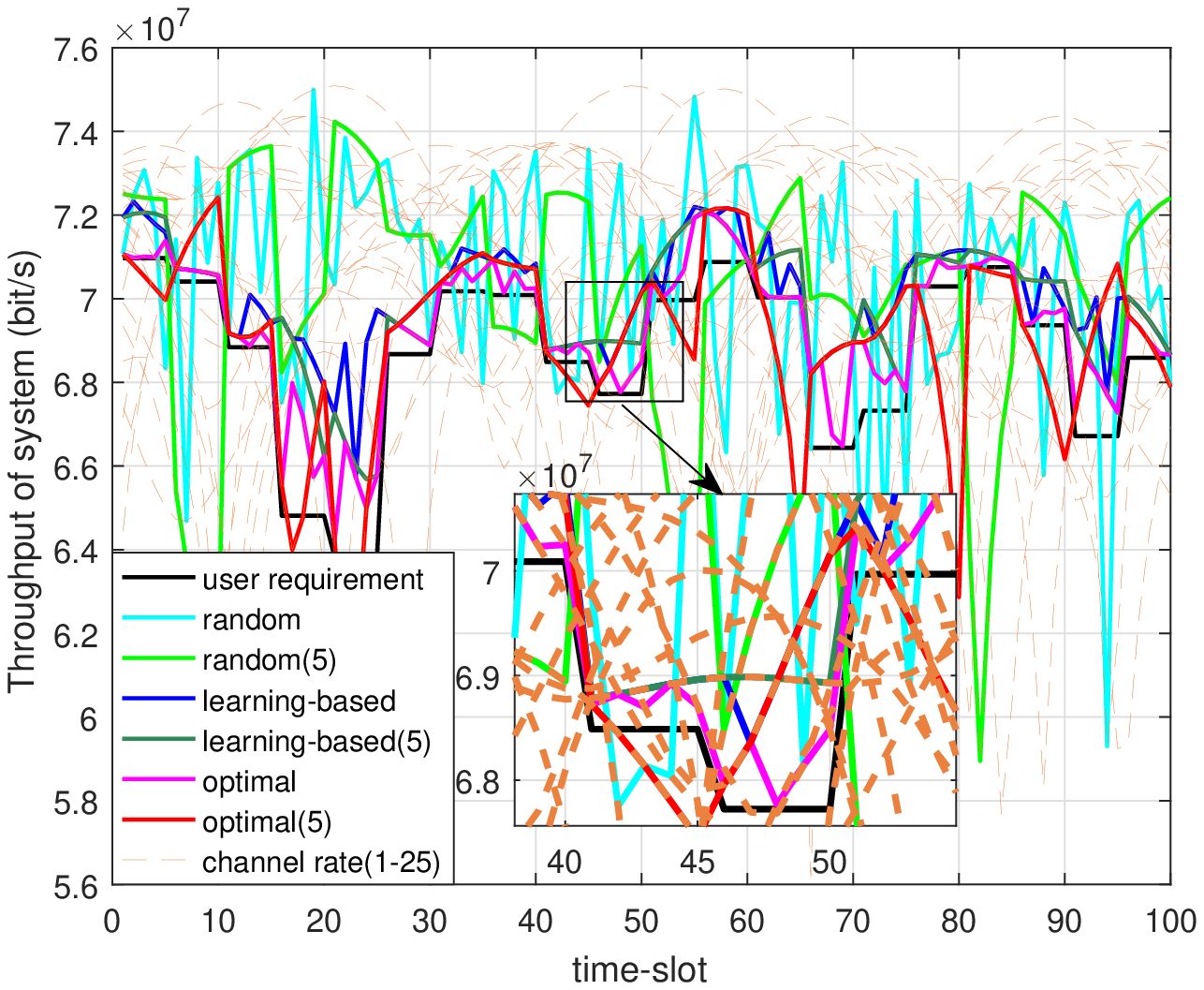}
	\caption{The single-user access in LSM.}\label{singel_user_access_no}
\end{figure}

Fig. \ref{singel_user_access_no} depicts the changes of the system throughput caused by different methods in \emph{one time-slot one decision} and \emph{$l$ time-slots one decision} over 100 time-slots, wherein $v=180$ km/h.
\begin{itemize}
  \item \emph{one time-slot one decision}: It can be seen that exhaustive search method (optimal) and the learning-based scheme can fully satisfy all user requirements, i.e., $\vartheta(t)=1$, and both of them are almost close to $R_{\mathrm{user},n}(t)$. The learning-based and the optimal methods are similar, but not identical, because the learning-based scheme considers the best performance for the next 5 time-slots.
  \item \emph{$l$ time-slots one decision}: In this case, the advantages of the learning-based scheme are fully revealed. The optimal method, due to the impact of rapid channel changes, is almost equivalent to the random method. Our learning-based scheme has negligible impact, still close to the optimal result of \emph{one time-slot one decision} scheme, which is quite impressive.
\end{itemize}
\begin{figure}[h]
	\centering{}\includegraphics[scale=0.55]{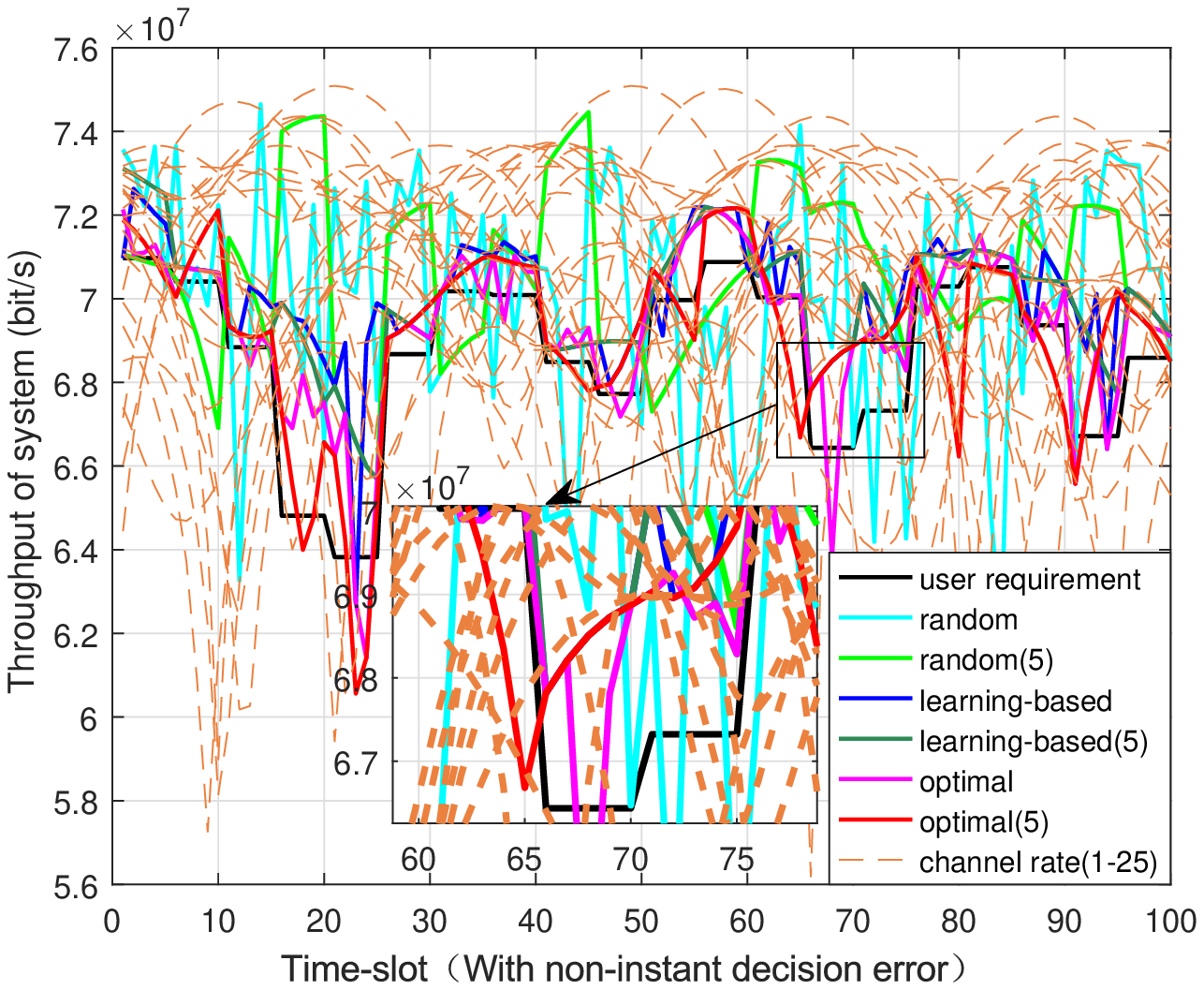}
	\caption{The single-user access in LSM with \emph{non-instant decision error}.}\label{singel_user_access_with}
\end{figure}

For the problem that traditional methods cannot handle \emph{non-instant decision error}, our learning-based scheme can greatly attenuate its impact, as shown in Fig. \ref{singel_user_access_with}. It is assumed that the information acquisition and processing delay $\varDelta t$ is 1 time-slot, which means that the strategy generated at time-slot $t$ will be executed at time-slot $t+1$. The \emph{service arrival rate} over $T$ time-slots is defined as
\begin{align}
\kappa(T)=\stackrel[t=1]{T}{\sum}(\varGamma(\vartheta(t)=1))/T.
\end{align}

In Fig. \ref{singel_user_access_with}, due to the influence of \emph{non-instant decision error}, the performance of the optimal method for \emph{one time-slot one decision} is greatly reduced, and its $\kappa(100)$ is only 0.83. The optimal method for \emph{$l$ time-slot one decision} violently vibrates as the channel changes, and $\kappa(100)$ is 0.64. In the learning-based scheme, $\kappa(100)=0.96$ in \emph{one time-slot one decision} and $\kappa(100)=1$ in \emph{$l$ time-slots one decision}. The fluctuation of the learning-based scheme, in both cases, is small, and the error between the rate of selected channel and the rate of user requirement satisfies criterion 1.
\subsubsection{Single-User Access in BSM}
\begin{figure*}[htb]
  \centering
  \subfigure[]{
    \includegraphics[width=3.1in]{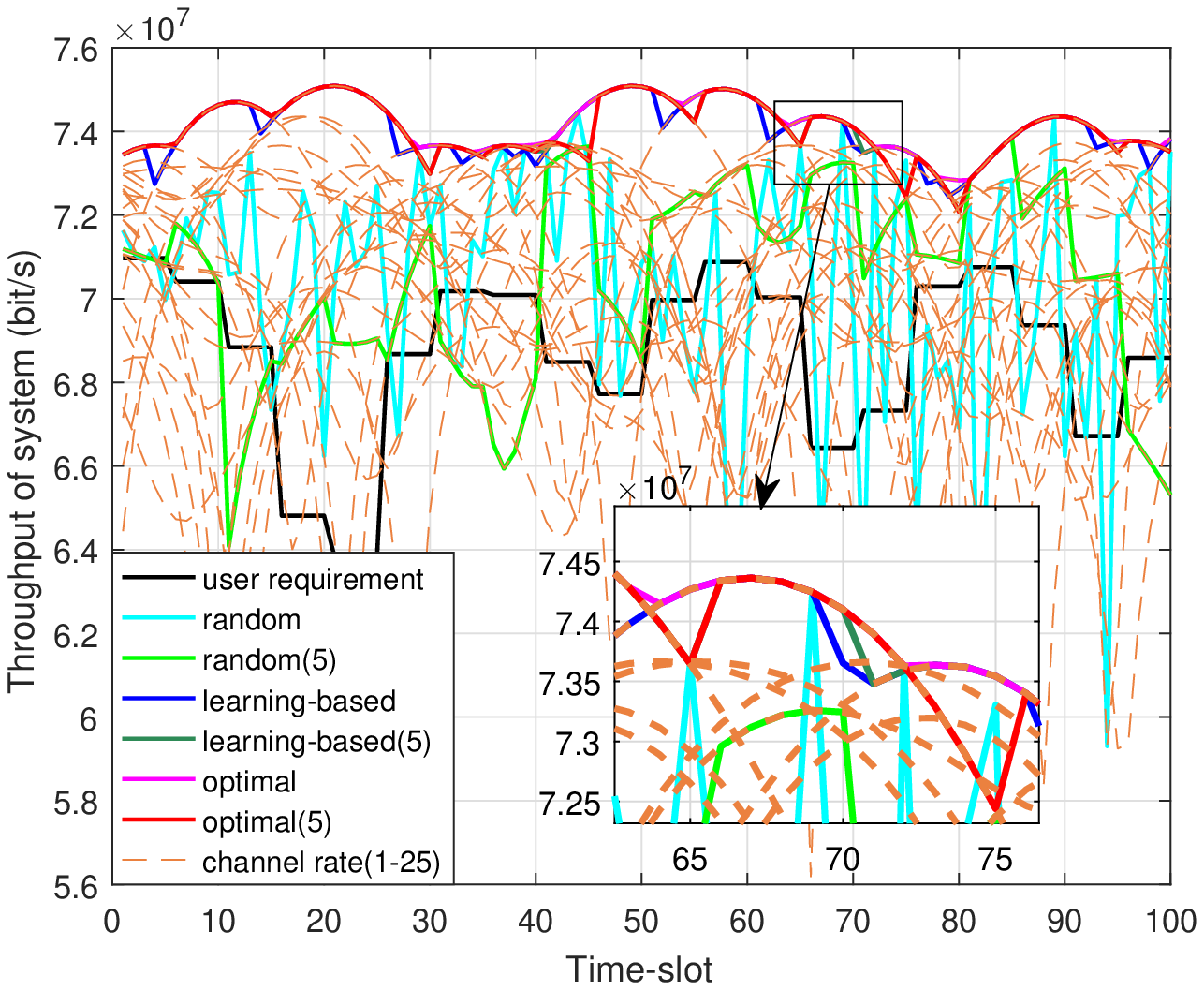}
  }
  \subfigure[]{
    \includegraphics[width=3.1in]{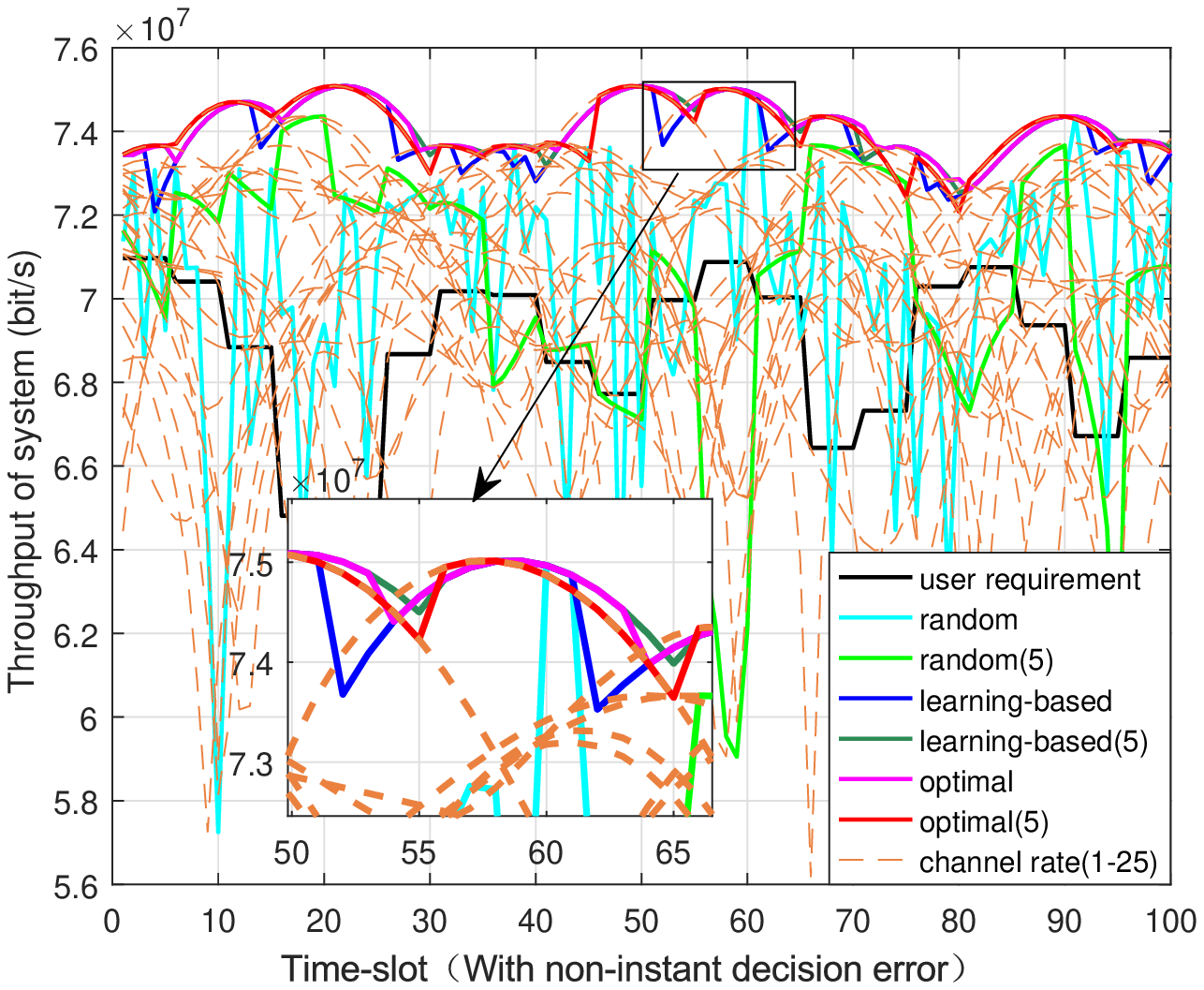}
  }
  \caption{The comparison of various methods for the single-user access in BSM.}
  \label{singel_user_access_buffer} 
\end{figure*}

Fig. \ref{singel_user_access_buffer}(a) and Fig. \ref{singel_user_access_buffer}(b) illustrate the performance of different methods without/with \emph{non-instant decision error}.
\begin{itemize}
  \item Without \emph{non-instant decision error}: In Fig. \ref{singel_user_access_buffer}(a), although the learning-based scheme is not dominant in the case of \emph{one time-slot one decision}, its performance is comparable to the optimal method. In the case of \emph{$l$ time-slots one decision}, the optimal method begins to deteriorate due to channel changes, but the learning-based scheme still maintains the highest performance in a stable manner.
  \item With \emph{non-instant decision error}: In Fig. \ref{singel_user_access_buffer}(b), the learning-based scheme, in addition to maintaining high performance, even performs better than the optimal method at some time-slots, revealing that it does reduce the impact of \emph{non-instant decision error}.
\end{itemize}

In Fig. \ref{singel_user_access_buffer}(a) and Fig. \ref{singel_user_access_buffer}(b), the throughput performance of the optimal method and the learning-based scheme is similar. This is because the channel change rates in the maximum point is small, and thus the changes of the channels have little impact on the optimal method. In fact, when the channel changes drastically, the optimal method will deteriorate sharply, which has been shown in detail in Fig. \ref{singel_user_access_no} and Fig. \ref{singel_user_access_with}.

\subsection{Multi-User Access}
\subsubsection{Multi-User Access in LSM}
\begin{figure*}[htb]
  \centering
  \subfigure[]{
    \includegraphics[width=3.1in]{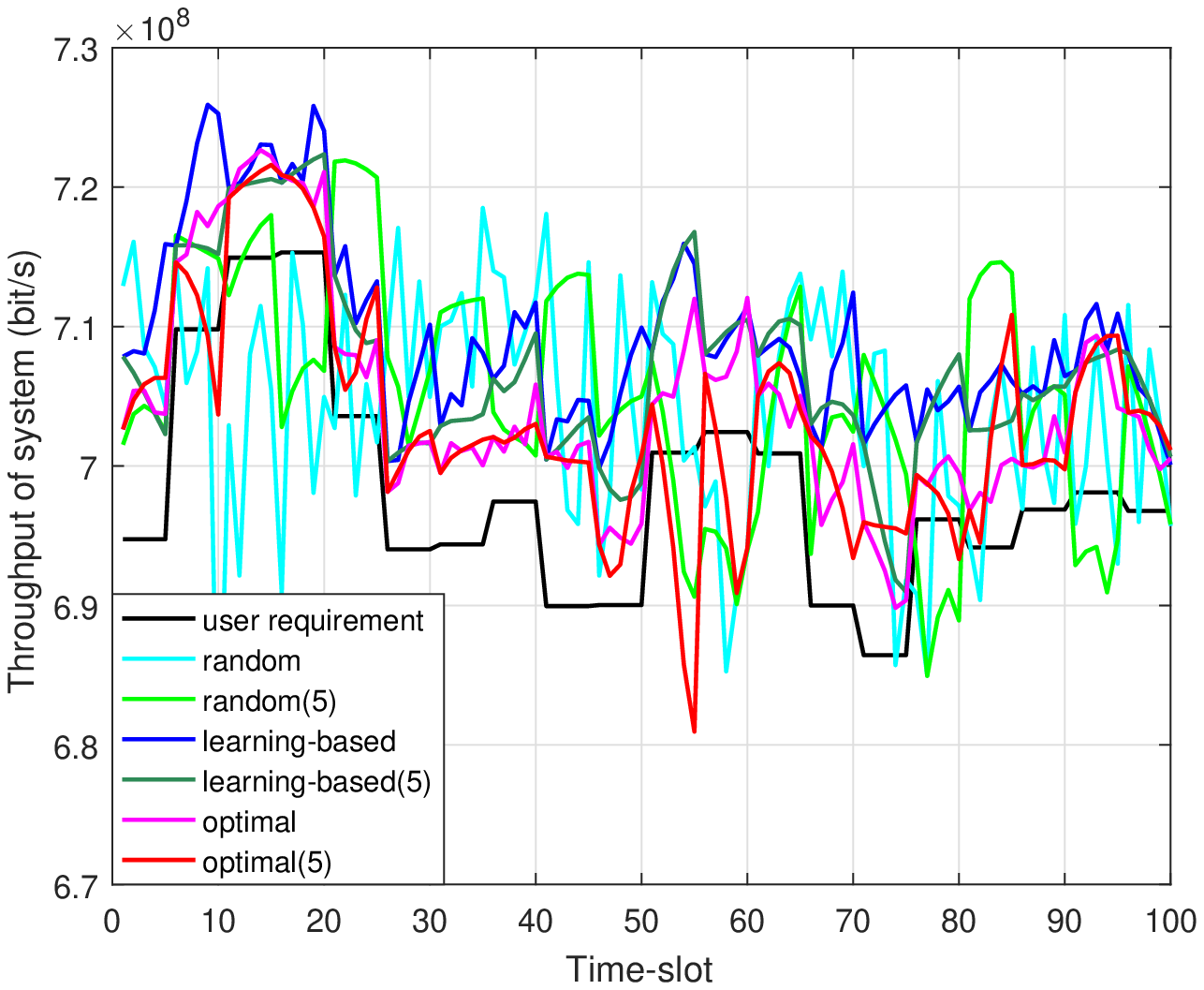}
  }
  \subfigure[]{
    \includegraphics[width=3.1in]{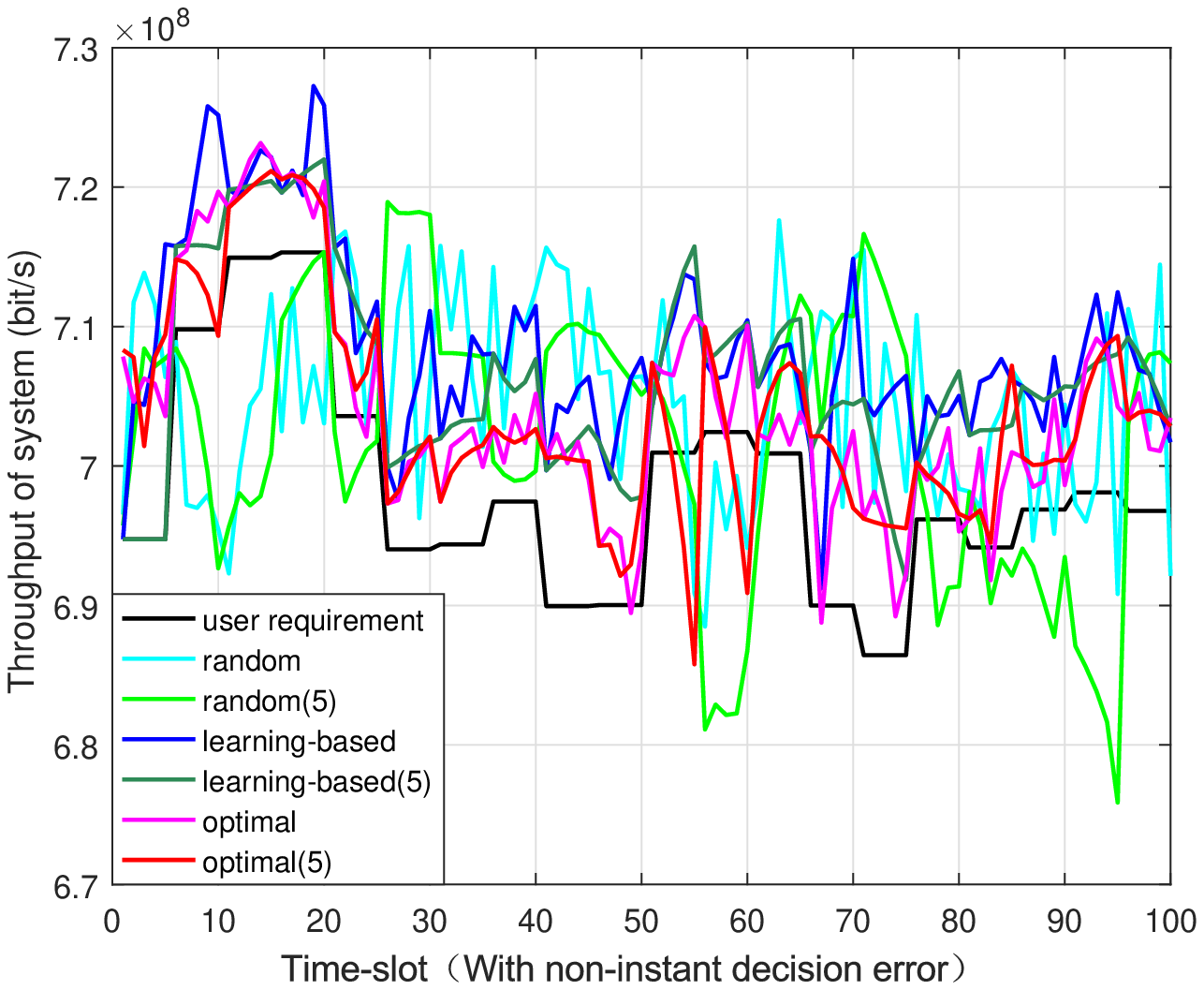}
  }
  \caption{The comparison of various methods for the multi-user access in LSM.}
  \label{multiple_user_access} 
\end{figure*}
Fig. \ref{multiple_user_access}(a) illustrates the performance of different methods without \emph{non-instant decision error}. It is worth noting that the value of the curve in Fig. \ref{multiple_user_access}(a) is the sum of the corresponding rates.
\begin{itemize}
  \item \emph{one time-slot one decision}: At each time-slot, the learning-based scheme can strictly satisfy criterion 1 and P-PQoS constraints of users.
  \item \emph{$l$ time-slots one decision}: The P-PQoS constraints cannot be met at some points, and $\kappa(100)=0.89$. For the learning-based scheme, the P-PQoS constraints can still be strictly satisfied, and $\kappa(100)=1$. Meanwhile, the performance of the learning-based scheme is superior, and there is no large fluctuation.
\end{itemize}

In Fig. \ref{multiple_user_access}(b), the advantages of the learning-based scheme are further highlighted. Due to the influence of \emph{non-instant decision error}, the optimal methods of \emph{one time-slot one decision} and \emph{$l$ time-slot one decision} cannot fully satisfy the P-PQoS constraints, and the values of $\kappa(100)$ are equal to 0.94 and 0.81. However, in the learning-based scheme, both the \emph{one time-slot one decision} and \emph{$l$ time-slot one decision} are affected to a small extent, and both of $\kappa(100)$ are 1.

\subsubsection{Multi-User Access in BSM}
\begin{figure*}[htb]
  \centering
  \subfigure[]{
    \includegraphics[width=3.1in]{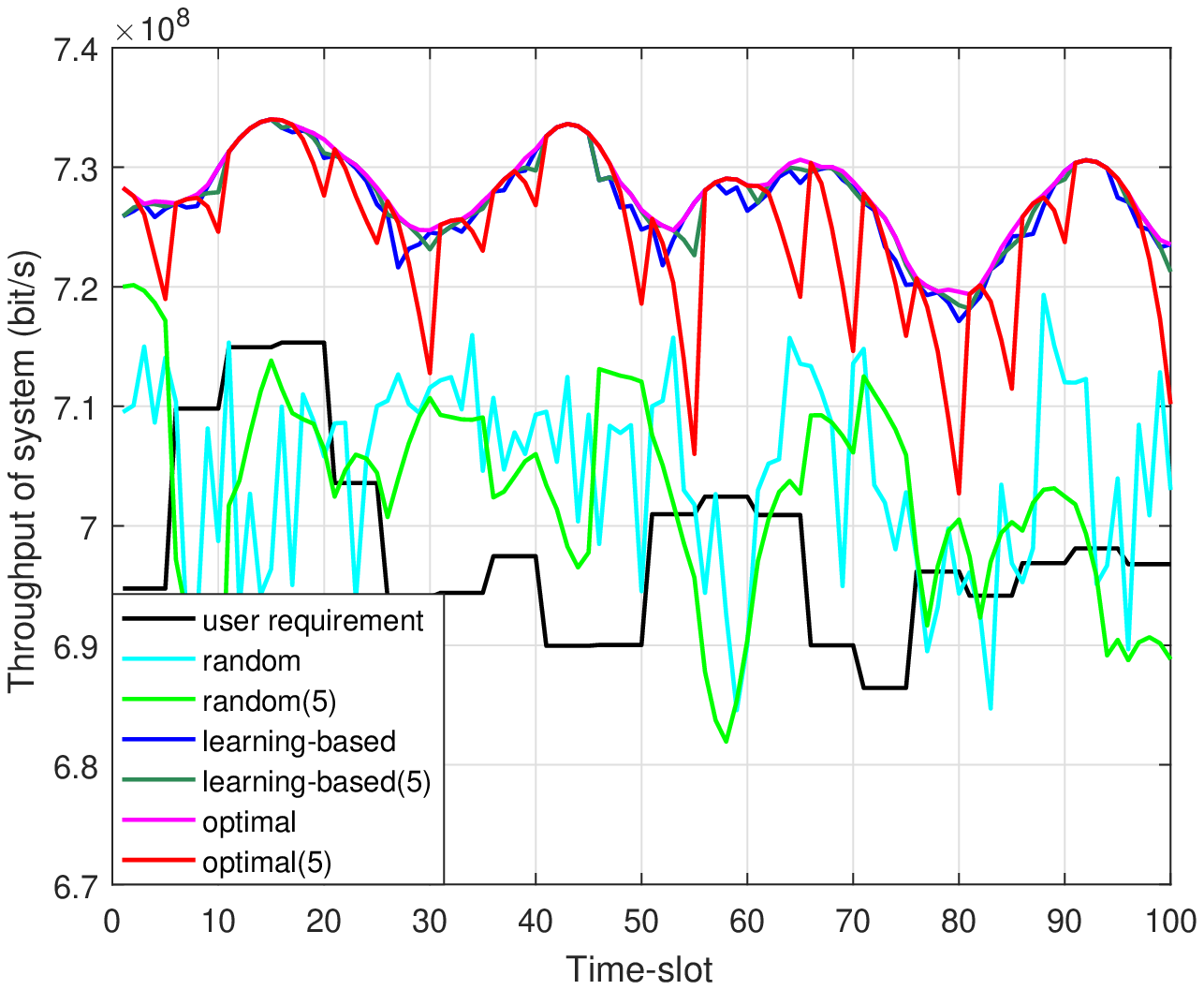}
  }
  \subfigure[]{
    \includegraphics[width=3.1in]{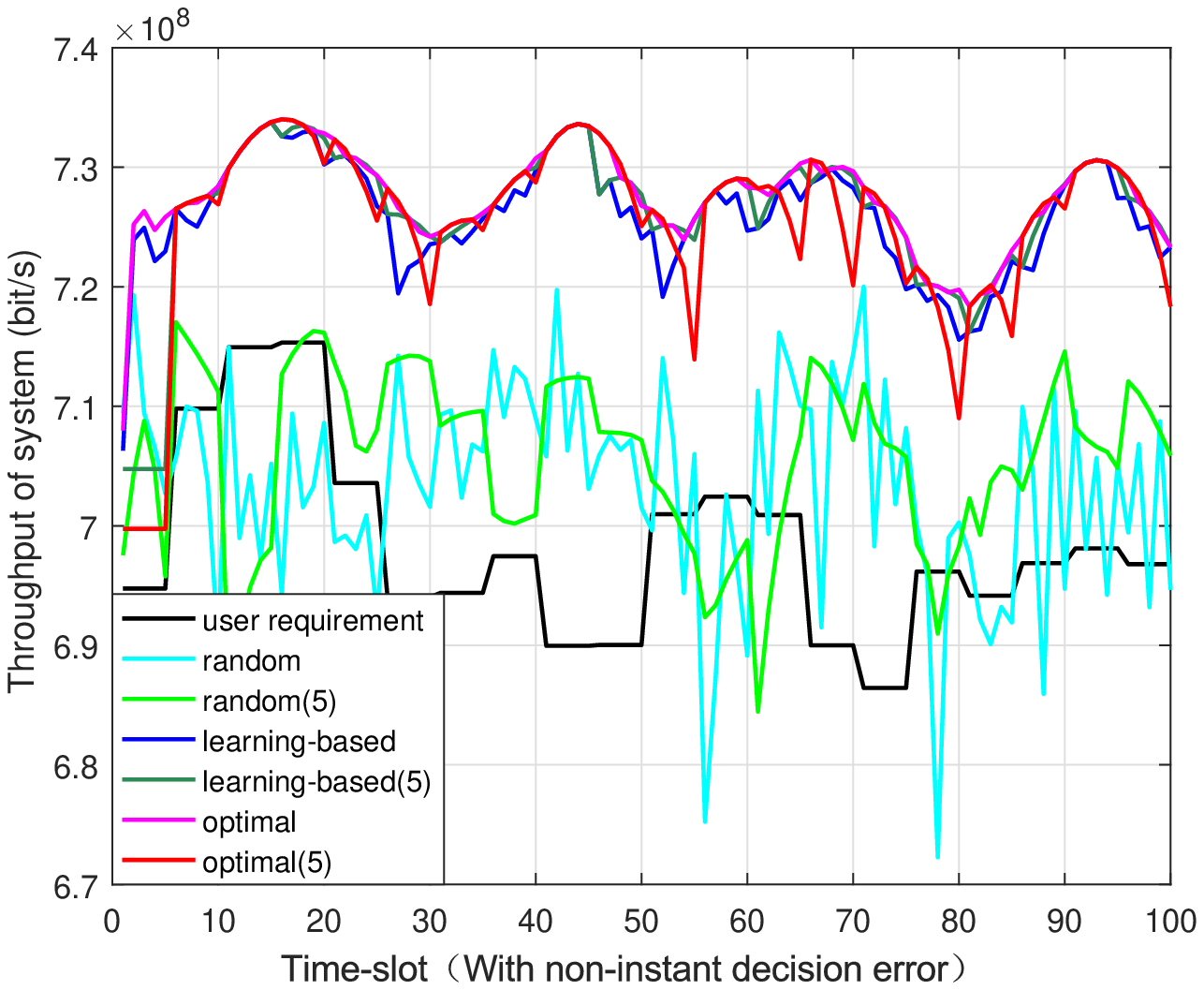}
  }
  \caption{The comparison of various methods for the multi-user access in BSM.}
  \label{multiple_user_access_buffer} 
\end{figure*}

The optimal performance of the learning-based scheme is close to the performance of \emph{one time-slot one decision} in Fig. \ref{multiple_user_access_buffer}(a). However, the optimal method does not perform well in \emph{$l$ time-slots one decision}. Fig. \ref{multiple_user_access_buffer}(b) also shows the impact of \emph{non-instant decision error} on the optimal method, exhibiting the superiority of our scheme.
\subsection{Service Stability Analysis}
Generally, it is irreconcilable to maintain the stability services and the adaptability for high-mobility systems. However, our approach can achieve excellent compromise.
The service stability is typically composed of three parts: the frequency of policy switching, the scale of policy switching and the fluctuation degree of adopted policy. The impacts caused by fluctuations in user requirements should be ignored, in order to characterize the stability of the service policy itself. Therefore, the \emph{service stability} is defined as
\begin{align}
S_{\mathrm{ta}}\!=\!&\left(\stackrel[i=1]{T/T_{\mathrm{one}}}{\sum}\mathbb{N}\left(\mathbf{A}(t\!+\!T_{\mathrm{one}}\!+\!i)\!-\!\mathbf{A}(t\!+\!i)\right)\right)\nonumber \\
&\!\times\!\frac{T}{T_{\mathrm{one}}}\times\left(\frac{1}{N}\stackrel[n=1]{N}{\sum}\left(F_{\mathrm{strategy},n}\!-\!F_{\mathrm{user},n}\right)\right),
\end{align}
where
\begin{align}
&F_{\mathrm{strategy},n}\!=\!E_{\mathrm{x}}\left(R_{\mathcal{A}_{n}(t)}(t),\cdots,R_{\mathcal{A}_{n}(t\!+\!T)}(t\!+\!T)\right)\times\frac{1}{T\!-\!1}\nonumber \\
&\!\times\!\stackrel[i=1]{T}{\sum}\left(R_{\mathcal{A}_{n}(t\!+\!i)}(t\!+\!i)\!-\!\frac{1}{T}\stackrel[i=1]{T}{\sum}\left(R_{\mathcal{A}_{n}(t\!+\!i)}(t\!+\!i)\right)\right),
\end{align}
and $\mathbb{N}\left(\mathbf{x}\right)$ represents the number of non-zero elements in vector $\mathbf{x}$. $T_{\mathrm{one}}$ is the switching interval between two policies. $E_{\mathrm{x}}\left(x\right)$ means the number of extreme values in sequence $x$. The fluctuation degree of user requirement $F_{\mathrm{user},n}$ is similar to $F_{\mathrm{strategy},n}$, and will not be detailed. Note that greater $S_{\mathrm{ta}}$ will result in the more unstable service.

\begin{figure*}[htb]
  \centering
  \subfigure[]{
    \includegraphics[width=2in]{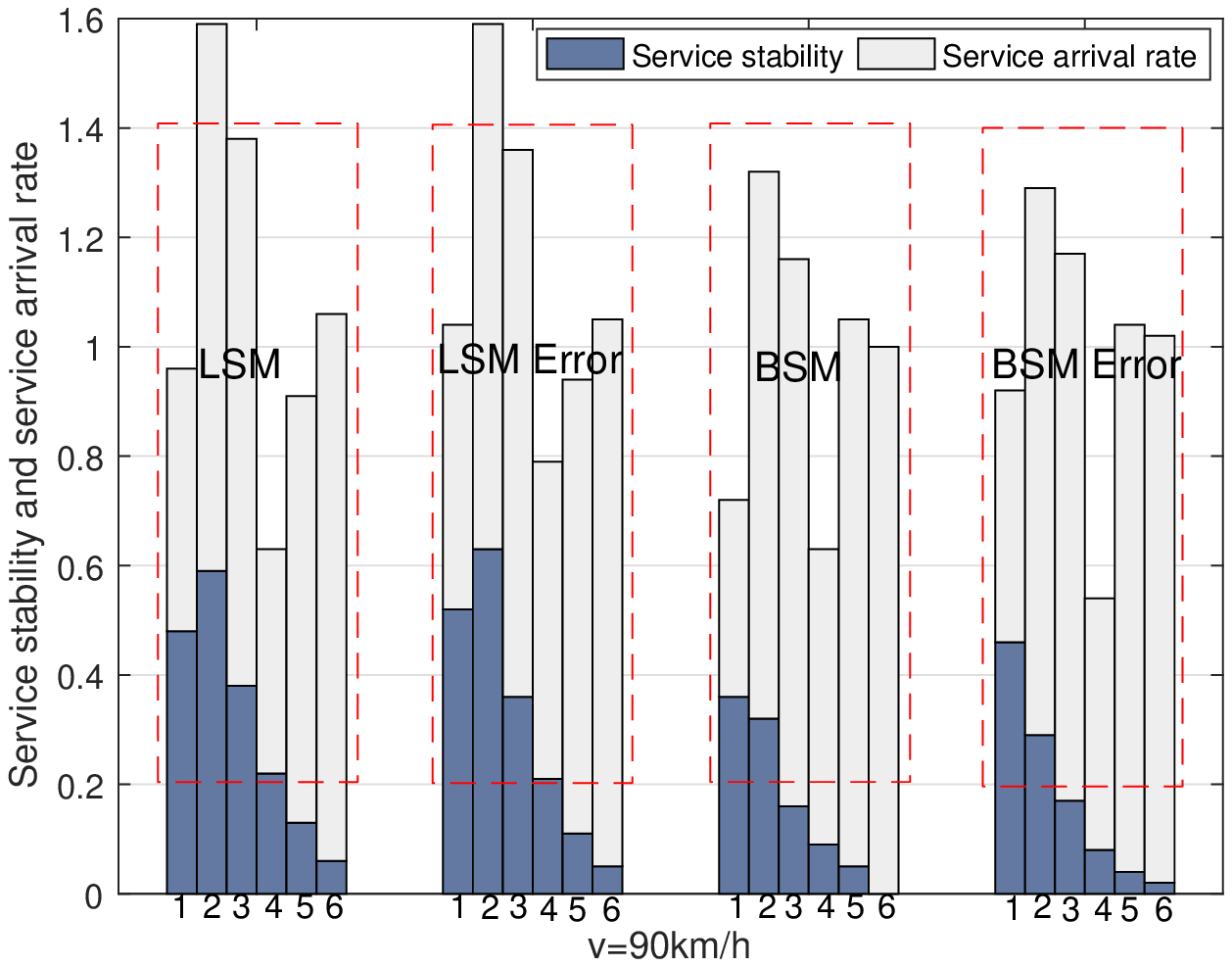}
  }
  \subfigure[]{
    \includegraphics[width=2in]{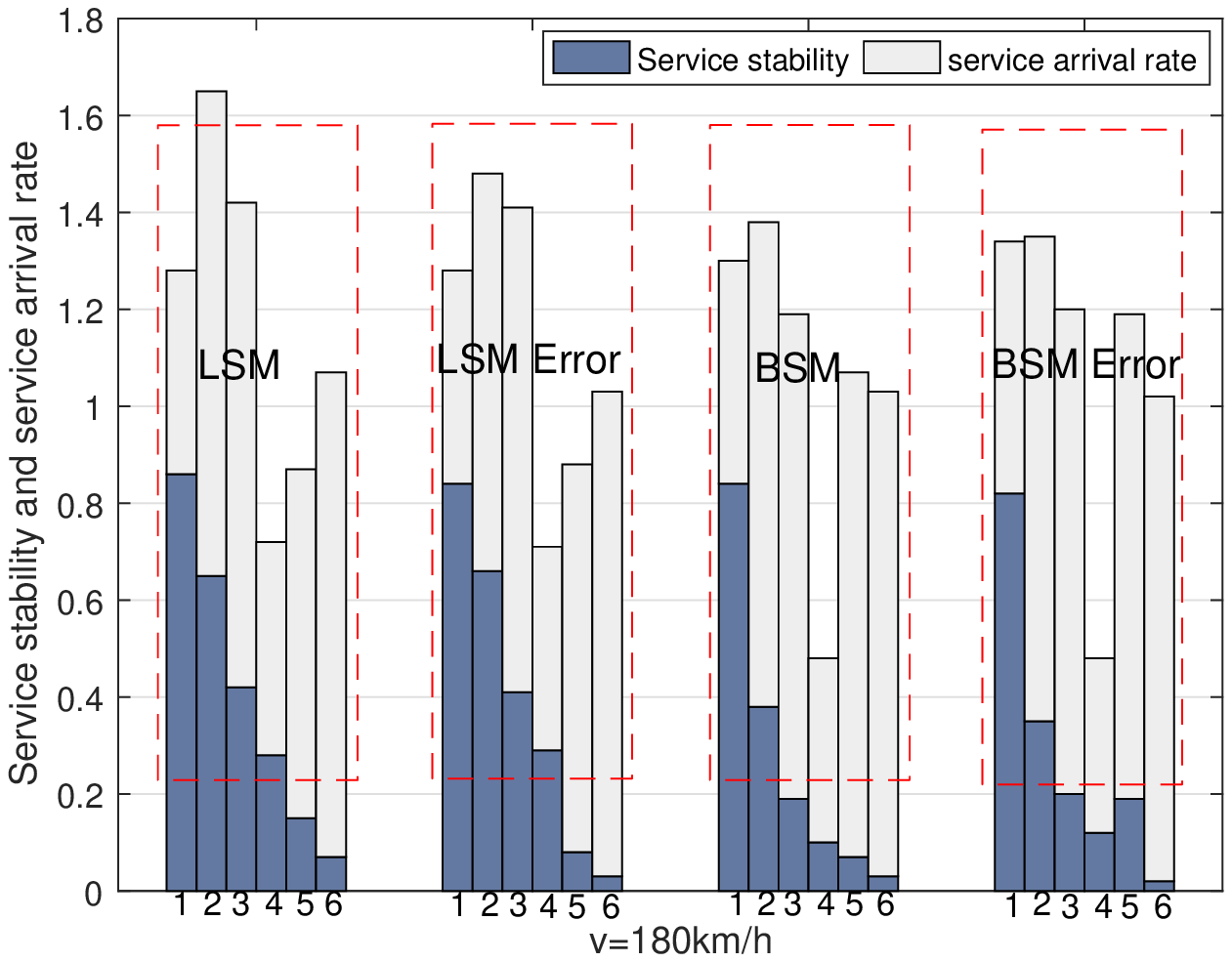}
  }
    \subfigure[]{
    \includegraphics[width=2in]{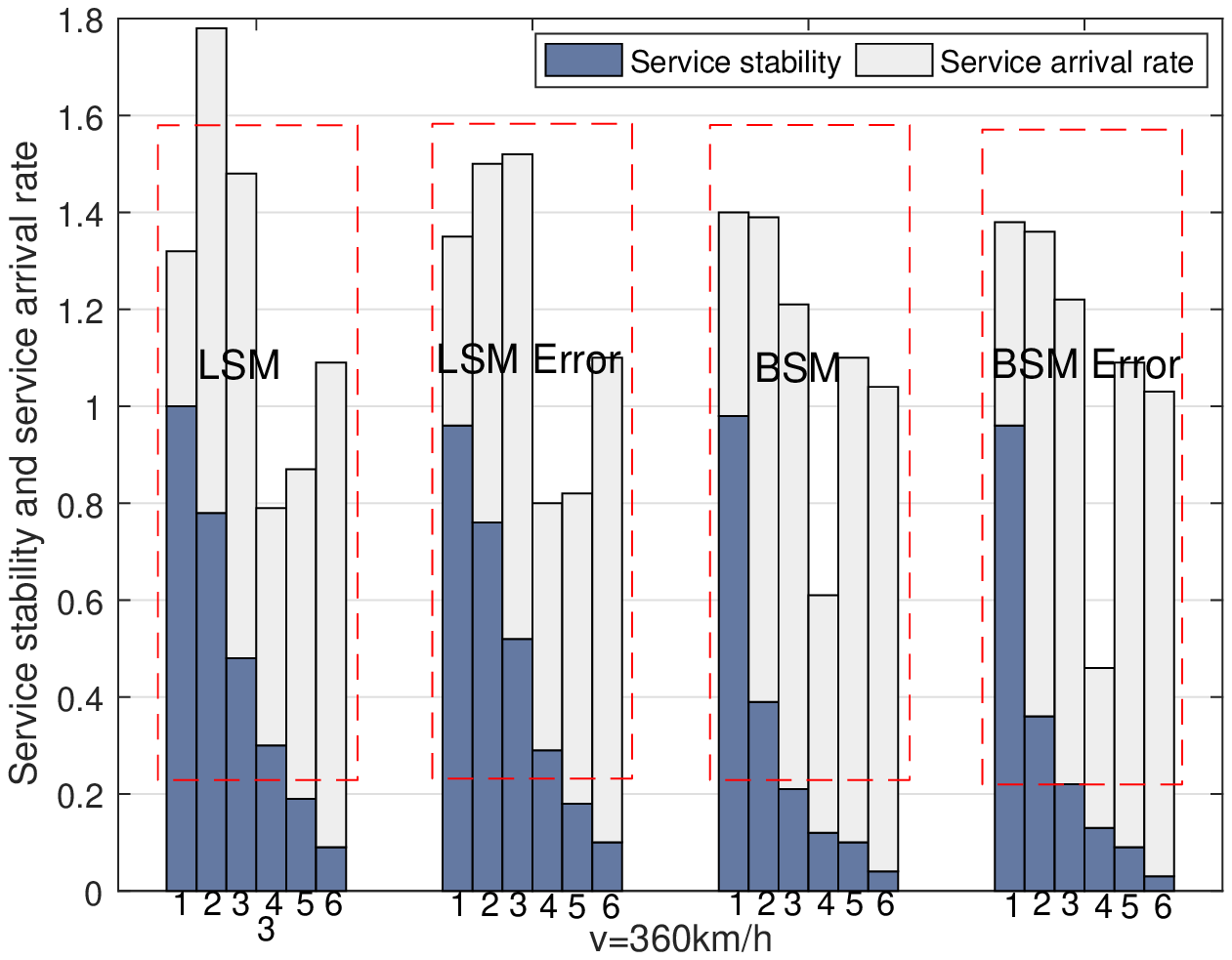}
  }
  \caption{The comparison of various methods for $\kappa(100)$ and the normalized $S_{\mathrm{ta}}$. Labels ``1,2,3,4,5,6'' mean different methods: ``random'', ``optimal'', ``learning-based'', ``random(5)'', ``optimal(5)'', ``learning-based(5)''. The ``LSM error'' means that \emph{non-instant decision error} is considered, and so are others.}
  \label{Stability} 
\end{figure*}

Fig. \ref{Stability} shows $S_{\mathrm{ta}}$ and $\kappa(100)$, and the value of $S_{\mathrm{ta}}$ has been normalized between 0 and 1. In \emph{one time-slot one decision}, the learning-based scheme is more stable than the optimal method, and the advantage is more pronounced in \emph{$l$ time-slots one decision}. In addition, the learning-based scheme can provide the highest $\kappa(100)$ in all situations.

\subsection{Impact of Prediction on the Model}
The above simulations have proven the advantages of the learning-based scheme. Inseparable from our P-DDPG algorithm, the proposed scheme can fully exploiting the characteristics of DRL and DMCA, and thus it can achieve such advantages.

\begin{figure*}[htb]
	\centering{}
    \includegraphics[width=0.95\textwidth]{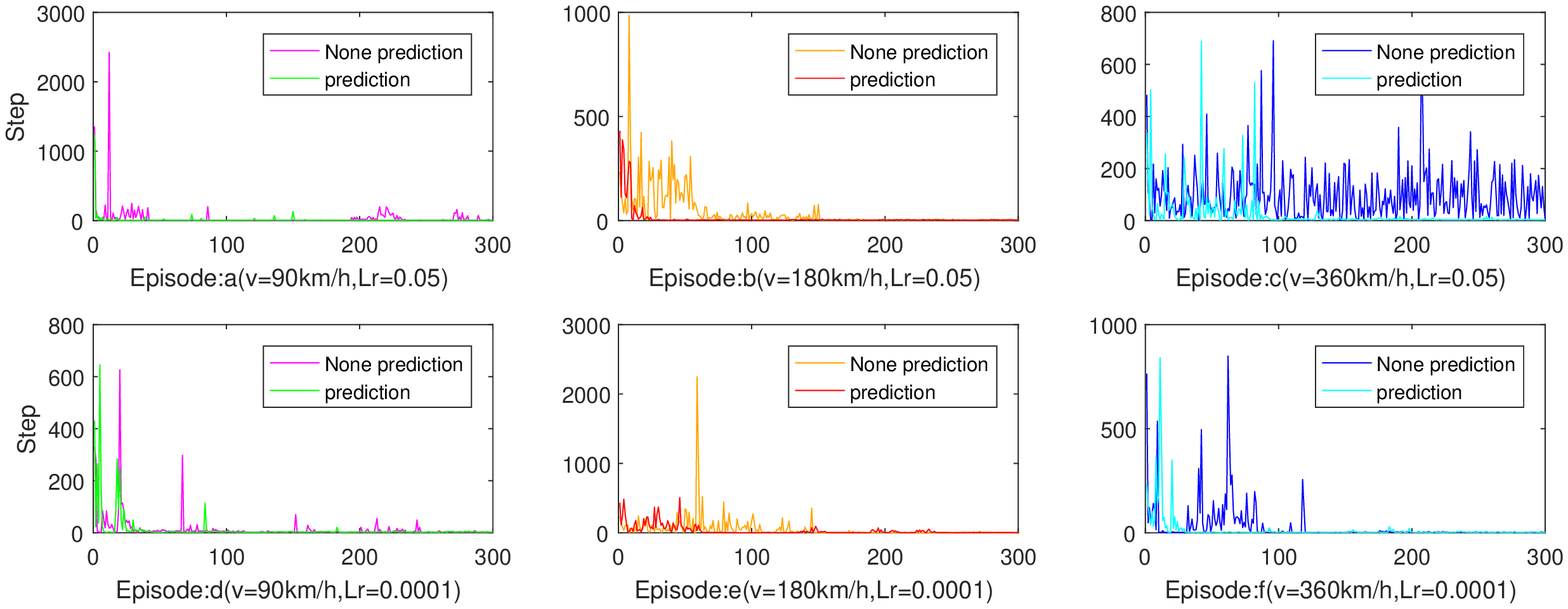}
	\caption{The effect of prediction on the convergence at different $v$ and Lr.}\label{prediction_yes_no}
\end{figure*}
The convergence performance of the learning-based DMCA for different learning rates (Lr) and $v$ are shown in Fig. \ref{prediction_yes_no}. The meaning of the abscissa \emph{episode} and the ordinate \emph{step} can be found in Algorithm 1 in detail. The smaller the \emph{step} is, the faster the learning-based scheme finds the optimal policy. Since $r_{\mathrm{dec}}=5$, the minimum value of \emph{step} is 5. \emph{step}=5 shows that the learning-based scheme can find the optimal solution in only one step.

We can see that the convergence rate with the prediction scheme is much faster than the one without prediction. In addition, the learning-based DMCA converges more slowly as $v$ increases. This is because the agent needs more explorations and learning in order to find more better reliable results as the channel fluctuations intensify. It should be noted that the agent even cannot converge in a short time in Fig. \ref{prediction_yes_no}(c), if the prediction is not added.

\section{Conlusion}
In this paper, we have deeply explored the application of DRL in DMCA according to the characteristics of 5G and beyond communication systems with the fast time-varying channels, and we also solved the DMCA problem by the learning-based scheme. The concept of P-PQoS is proposed to portray the service delay differences in \emph{subjective experience}. The proposed CPM with the high-accuracy rate has proven the channel predictability under the specific scenarios, and may provide new inspirations for channel estimation/modeling. The real channel data-based simulation results have validated that the performance of the learning-based scheme approaches the exhaustive search method when making decisions at each time-slot, and is superior to the exhaustive search method when making decisions every few time-slots. Meanwhile, the scheme greatly weakens the impact of  \emph{non-instant decision error} while satisfying P-PQoS requirements. Our learning-based DMCA scheme is highly scalable and flexible, and can be easily migrated to other scenarios. The design ideas and methods are expected to unfold into a new technical paradigm.

\bibliographystyle{IEEEtran}

\begin{IEEEbiography}[{\includegraphics[width=1in,height=1.25in,clip,keepaspectratio]{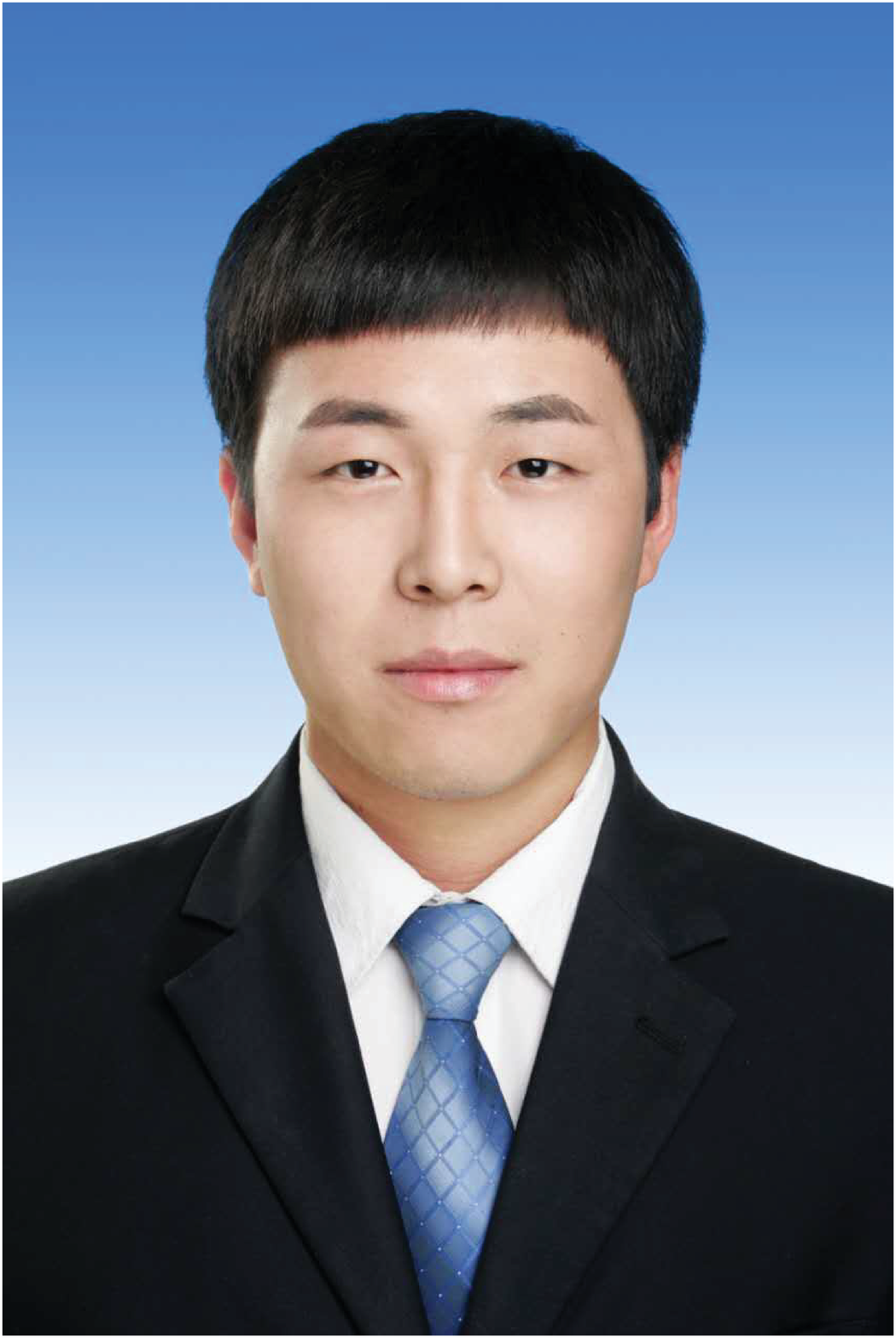}}]{Shaoyang Wang}
(S'18) received the B.E. degree in electronic information science and technology from Shandong University, Weihai, China, in 2017. He is currently pursuing the Ph.D. degree with the School of Information and Communication Engineering, Beijing University of Posts and Telecommunications (BUPT), Beijing, China. His current research interests include non-orthogonal multiple access access, network function virtualization, and intelligent wireless resource management.
\end{IEEEbiography}
\begin{IEEEbiography}[{\includegraphics[width=1in,height=1.25in,clip,keepaspectratio]{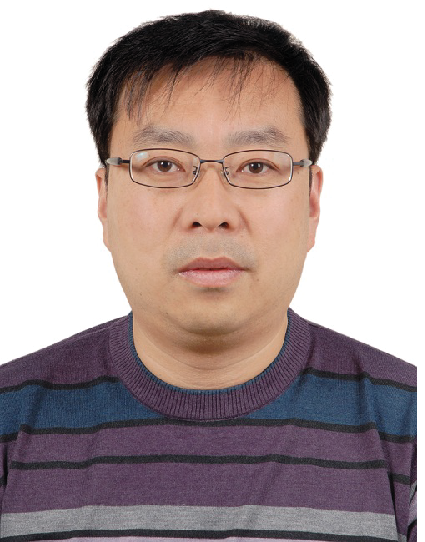}}]{Tiejun Lv}
(M'08-SM'12) received the M.S. and Ph.D. degrees in electronic engineering from the University of Electronic Science and Technology of China (UESTC), Chengdu, China, in 1997 and 2000, respectively. From January 2001 to January 2003, he was a Postdoctoral Fellow with Tsinghua University, Beijing, China. In 2005, he was promoted to a Full Professor with the School of Information and Communication Engineering, Beijing University of Posts and Telecommunications (BUPT). From September 2008 to March 2009, he was a Visiting Professor with the Department of Electrical Engineering, Stanford University, Stanford, CA, USA. He is the author of 2 books, more than 70 published IEEE journal papers and 180 conference papers on the physical layer of wireless mobile communications. His current research interests include signal processing, communications theory and networking. He was the recipient of the Program for New Century Excellent Talents in University Award from the Ministry of Education, China, in 2006. He received the Nature Science Award in the Ministry of Education of China for the hierarchical cooperative communication theory and technologies in 2015.
\end{IEEEbiography}
\begin{IEEEbiography}[{\includegraphics[width=1in,height=1.25in,clip,keepaspectratio]{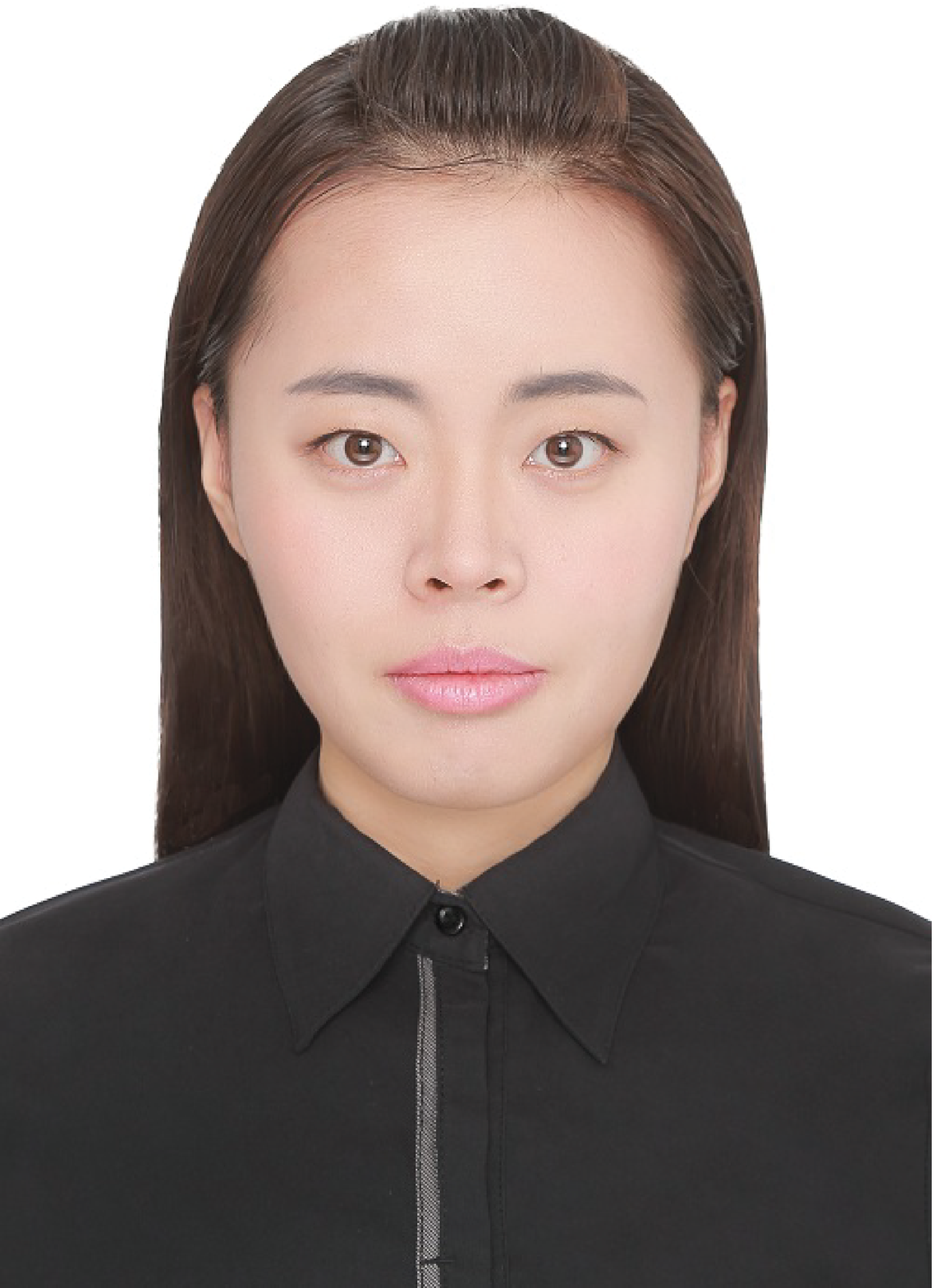}}]{Xuewei Zhang}
(S'18) received the B.E. degree in communication engineering from Tianjin Polytechnic University, Tianjin, China, in 2015. She is currently pursuing the Ph.D. degree with the School of Information and Communication Engineering, Beijing University of Posts and Telecommunications (BUPT), Beijing, China. Her research interests include multicast beamforming, wireless caching and resource allocation.
\end{IEEEbiography}
\begin{IEEEbiography}[{\includegraphics[width=1in,height=1.25in,clip,keepaspectratio]{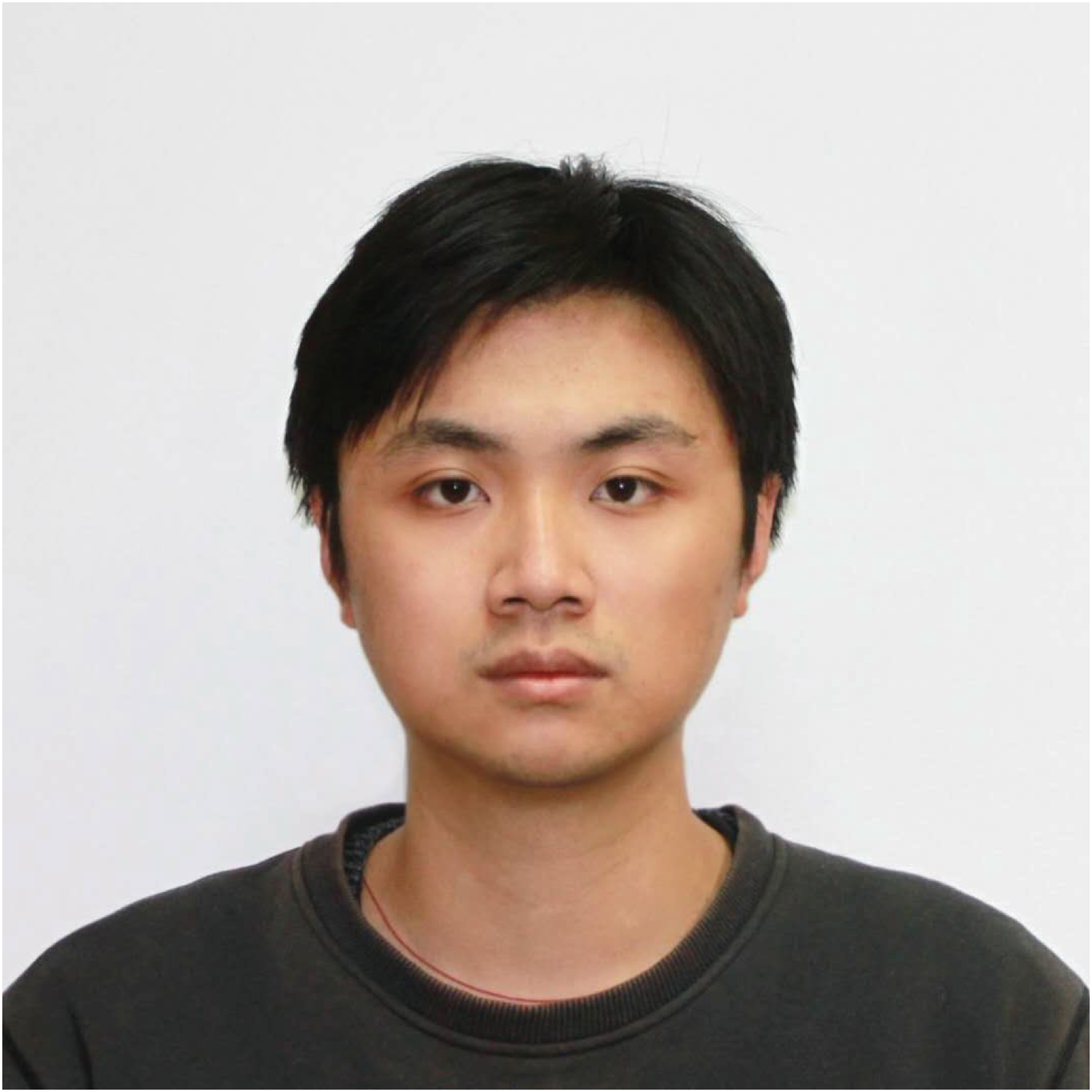}}]{Zhipeng Lin}
(S'17) is currently pursuing the dual Ph.D. degrees in communication and information engineering with the School of Information and Communication Engineering, Beijing University of Posts and Telecommunications, Beijing, China, and the School of Electrical and Data Engineering, University of Technology of Sydney, Sydney, NSW, Australia. His current research interests include millimeter-wave communication, massive MIMO, hybrid beamforming, wireless localization, and tensor processing.
\end{IEEEbiography}
\begin{IEEEbiography}[{\includegraphics[width=1in,height=1.25in,clip,keepaspectratio]{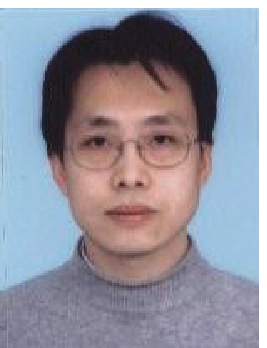}}]{Pingmu Huang}
received the M.S. degrees from Xi¡¯an Jiaotong University, Xian, China, in 1996 and received Ph.D. degree of Signal and Information Processing from Beijing University of Posts and Telecommunications (BUPT), Beijing, China, in 2009. He is now a lecturer with the School of Information and Communication Engineering, BUPT. His current research interests include signal processing and pattern recognition. He published several journal papers and conference papers on signal processing and pattern recognition.
\end{IEEEbiography}
\end{document}